\documentclass{ws-procs961x669}            
\usepackage{slashed} 
\usepackage{subfigure} 
\usepackage{multirow}
\usepackage{hyperref}
\begin{document}
\title{Leptophilic Dark Matter at Linear Colliders}

\author{P. S. Bhupal Dev}

\address{Department of Physics and McDonnell Center for the Space Sciences, Washington University, 
St.\,Louis, MO 63130, USA\\
E-mail: bdev@wustl.edu}

\begin{abstract}
We discuss model-independent collider constraints on the effective couplings of leptophilic dark matter (LDM), considering its production at a future electron-positron linear collider, with both polarized and unpolarized beam options, in the mono-photon and mono-$Z$ channels. We show that the future collider constraints are largely complementary to the direct and indirect detection limits on LDM, and can potentially provide the best-ever LDM sensitivity in the low-mass regime (below $\sim$ 10 GeV).  
\end{abstract}


\bodymatter

\section{Introduction}
\label{sec:intro}
Many of the existing experimental constraints on dark matter (DM) crucially rely on the DM interactions with nucleons, and therefore, can be largely weakened if the DM predominantly interacts with the Standard Model (SM) leptons, but not quarks at tree-level. Such {\it leptophilic} DM (LDM) could  arise naturally in many beyond the Standard Model (BSM) scenarios~\cite{Krauss:2002px, Baltz:2002we, Ma:2006km, Hambye:2006zn, Bernabei:2007gr, Cirelli:2008pk, Chen:2008dh, Bi:2009md, Ibarra:2009bm, Dev:2013hka, Chang:2014tea, Agrawal:2014ufa, Bell:2014tta, Freitas:2014jla,Cao:2014cda, Lu:2016ups, Duan:2017pkq,Madge:2018gfl,Junius:2019dci, Ghosh:2020fdc,  Chakraborti:2020zxt, Horigome:2021qof}, some of which could even explain various experimental anomalies, such as the muon anomalous magnetic moment
~\cite{Abi:2021ojo}, DAMA/LIBRA annual modulation~\cite{Bernabei:2020mon}, 
anomalous cosmic ray positron excess~\cite{Abdollahi:2017nat, DAMPE:2017fbg, Adriani:2018ktz, AMS:2021nhj}, 
the galactic center gamma-ray excess~\cite{TheFermi-LAT:2015kwa}, and XENON1T electron excess~\cite{XENON:2020rca}. Dedicated searches for LDM in direct detection~\cite{XENON100:2015tol, XENON:2019gfn, LZ:2021xov} and beam dump~\cite{Chen:2018vkr, Marsicano:2018vin} experiments have also been discussed. 
 
 In this proceeding based on Ref.~\cite{Kundu:2021cmo}, we focus on the LDM searches at lepton colliders, which are complementary to the direct and indirect detection searches. We adopt an effective field theory (EFT) approach, which has been widely used in the context of collider searches for DM following the early works of Refs.~\cite{Kopp:2009et, Beltran:2010ww,Goodman:2010yf,Bai:2010hh, Goodman:2010ku, Fox:2011fx,Fox:2011pm, Rajaraman:2011wf,Chae:2012bq}.  The same interactions responsible for DM pair-annihilation in the early universe leading to their thermal freeze-out guarantee their direct production at colliders, as long as kinematically allowed. This will give a characteristic mono-$X$ signature, where the large missing transverse momentum carried away by the DM pair is balanced by a visible sector particle $X$ (which can be either a photon, jet, $W$, $Z$, or Higgs, depending on the model) emitted from an initial, intermediate or final state (see Refs.~\cite{Kahlhoefer:2017dnp, Penning:2017tmb} for reviews). Specifically, the mono-jet signature has become emblematic for LHC DM searches~\cite{CMS:2014jvv, ATLAS:2015qlt, CMS:2017zts, ATLAS:2021kxv}. However, for an LDM with loop-suppressed interactions to the SM quarks, the hadron colliders like the LHC are not expected to provide a better limit than the existing constraints from indirect searches, such as from AMS-02~\cite{Cavasonza:2016qem, John:2021ugy}, at least within the EFT framework with contact interactions. 
 
 On the other hand, lepton colliders provide an ideal testing ground for the direct production of LDM and its subsequent detection via either mono-photon~\cite{DELPHI:2003dlq, Birkedal:2004xn, Fox:2008kb,Konar:2009ae, Fox:2011fx,Bartels:2012ex, Dreiner:2012xm,Chae:2012bq, Liu:2019ogn,Habermehl:2020njb, Kalinowski:2021tyr, Barman:2021hhg} or mono-$Z$~\cite{Wan:2014rhl, Yu:2014ula, Dutta:2017ljq, Grzadkowski:2020frj} signatures. We go beyond the existing literature and perform a comprehensive and comparative study of both mono-photon and mono-$Z$ signatures of LDM at future $e^+e^-$ colliders in a  model-independent, EFT approach~\cite{Kundu:2021cmo}. Our analysis is generically applicable to all future $e^+e^-$ colliders, such as  the ILC~\cite{Bambade:2019fyw}, CLIC~\cite{CLIC:2016zwp}, CEPC~\cite{CEPCStudyGroup:2018ghi} and FCC-ee~\cite{FCC:2018evy}, but for concreteness, we have taken the $\sqrt s=1$ TeV ILC as our case study for numerical simulations. We also assume the DM to be fermionic and limit ourselves to the dimension-6 operators, but taking into consideration all possible dimension-6 operators of scalar-pseudoscalar (S-P), vector-axialvector (V-A) and tensor-axialtensor (T-AT) type as applicable for the most general DM-electron coupling. 
 Within the minimal EFT approach, the only relevant degrees of freedom in our analysis are the DM mass and an effective cut-off scale $\Lambda$ which determines the strength of the four-Fermi operators. This enables us to derive model-independent ILC sensitivities on LDM in the $(m_\chi, \Lambda)$ plane in both mono-photon and mono-$Z$ (leptonic and hadronic) channels, after taking into account all relevant backgrounds and systematic uncertainties.  We consider both unpolarized and polarized beam options~\cite{Bambade:2019fyw, Barklow:2015tja}, and find that with the proper choice of polarizations for the $e^-$ and $e^+$ beams (which depends on the operator type), the DM sensitivities could be significantly enhanced. 

\section{Effective operators}\label{sec:2}

Our primary assumptions are (i) the DM particle $\chi$ couples directly only to the SM leptons but not to the quarks (hence leptophilic), and (ii) the energy scale of the associated new physics is large compared to the collider energies under consideration, thus allowing us to integrate out the heavy mediators and parametrize the DM-SM interactions using effective higher-dimensional operators.    
For concreteness, we assume that the DM particles are Dirac fermions, and therefore, the leading order DM-SM interactions are the dimension-six four-Fermi interactions, with the most-general
effective Lagrangian given by~\cite{Kopp:2009et} 
\begin{align}
    \mathcal{L}_{\rm eff}  =  \frac{1}{\Lambda^2}\sum_j\left(\overline{\chi}\Gamma^{j}_{\chi}\chi\right)\left(\overline{\ell}\Gamma^{j}_{\ell}\ell\right) \, , 
    \label{eq:EFT}
\end{align}
 where $\Lambda$ is the cut-off scale for the EFT description and the index $j$ corresponds to different Lorentz structures, as shown below. Since our main focus is on $e^+e^-$ colliders, we will just set $\ell=e$ in Eq.~\eqref{eq:EFT} and assume this to be the only leading-order coupling, but our discussion below could be easily extended to other cases, e.g. future muon colliders~\cite{Delahaye:2019omf} by setting $\ell=\mu$. 
 
 A complete set of Lorentz-invariant operators consists of scalar (S),  pseudo-scalar (P), vector (V), axial-vector (A), tensor (T) and axial-tensor (AT) currents. We classify them as follows:%
   \begin{align}
 &\text{S-P type}: & &
   \Gamma_{\chi}  =  c^{\chi}_{S}+i c^{\chi}_{P} \gamma_5 \, ,  & & \Gamma_{e} = c^{e}_{S}+i c^{e}_{P} \gamma_5 \, , \nonumber \\
   &\text{V-A type}: & &
   \Gamma_{\chi}^{\mu} =  \left( c^{\chi}_{V}+ c^{\chi}_{A} \gamma_5 \right) \gamma^{\mu} \, ,  & &
   \Gamma_{e\mu} = \left( c^{e}_{V}+ c^{e}_{A} \gamma_5 \right)\gamma_{\mu}  \, , \nonumber \\
   &\text{T-AT type}: & &
   \Gamma_{\chi}^{\mu \nu} = \left( c^{\chi}_{T}+i c^{\chi}_{AT} \gamma_5 \right) \sigma^{\mu \nu} \, , & & \Gamma_{e \mu \nu} = \sigma_{\mu \nu} \, ,
    \label{eq:operator}
   \end{align}
   where $\sigma^{\mu \nu}=\frac{i}{2}[\gamma^\mu,\gamma^\nu] $ is the spin tensor and $c_j^{\chi,e}$ are dimensionless, real couplings. 
   For simplicity, in Eq.~\eqref{eq:EFT} we have used a common cut-off scale $\Lambda$ for all Lorentz structures. Furthermore, in our subsequent numerical analysis, we will consider one type of operator at a time, by setting the corresponding couplings $c_j^{\chi,e}=1$ without loss of generality and all other couplings equal to zero, unless otherwise specified. For instance, setting $c_S^\chi=c_P^\chi=c_S^e=c_P^e=1$ and all other couplings equal to zero gives us the (S+P)-type operator, which we will simply refer to as the {\bf SP}-type in the following discussion. Similarly, we will denote the  $c_V^\chi=c_A^\chi=c_V^e=c_A^e=1$ case simply as the {\bf VA}-type, and  $c_T^\chi=c_{AT}^\chi=1$ as the {\bf TAT}-type for presenting our numerical results in the $(m_\chi,\Lambda)$ plane. For other choices of the couplings, our results for the sensitivity on $\Lambda$ can be easily scaled accordingly.    
   
We will impose a theoretical limit of $\Lambda>\sqrt s$ for the EFT validity. For relatively larger DM mass, we must also have $\Lambda>2m_\chi$ in order to describe DM pair annihilation by the EFT. In fact, using $\Lambda=2m_\chi$ induces 100\% error in the EFT prediction for $s$-channel UV completions. Therefore, we will use $\Lambda>{\rm max}\{\sqrt s, 3m_\chi\}$ as a conservative lower bound~\cite{Matsumoto:2016hbs} to ensure the validity of our EFT approach.




\section{Mono-photon channel} \label{sec:3}


For the mono-photon signal  $e^+e^-\to\chi\overline{\chi}\gamma$, the $\chi$'s will contribute to the missing transverse energy at the detector. The dominant irreducible SM background to this process comes from neutrino pair production with an associated ISR photon, i.e. $e^+e^-\to\nu\overline{\nu}\gamma$. Since neutrinos are practically indistinguishable from DMs on an event-by-event basis, the majority of this background survives the event selection cuts. However, as we will show later, this background is highly polarization-dependent, and therefore, can be significantly reduced by the proper choice of polarized beams, without affecting the signal much. 

Apart from the neutrino background, any SM process with a single photon in the final state can contribute to the total background if all other visible particles escape detected. The SM processes containing either jets or charged particles are relatively easy to distinguish from a DM event, so their contribution to the total background is negligible~\cite{Bartels:2012ex}. The only exception is the \emph{Bhabha scattering} process associated with an extra photon (either from initial or final state radiation), i.e., $e^+e^-\to e^+e^-\gamma$, which has a large cross section, is polarization-independent, and can significantly contribute to the total background whenever the final-state electrons and positrons go undetected, e.g. along beam pipes. In our following analysis, we consider both neutrino and radiative Bhabha backgrounds.  

\subsection{Cross-sections}\label{sec:3.1} 
The cross-sections for the mono-photon signal $e^+e^-\to \chi\overline{\chi}\gamma$ and the radiative neutrino background $e^+e^-\to \nu\overline\nu\gamma$ at $\sqrt s=1$ TeV ILC are estimated using \texttt{CalcHEP}~\cite{Belyaev:2012qa} with proper implementation of ISR and beamsstrahlung effects, which significantly affect the width and position of the neutrino $Z$-resonance. For this purpose, the EFT Lagrangian~\eqref{eq:EFT} is implemented in \texttt{FeynRules}~\cite{Alloul:2013bka} to generate the CHO library required for \texttt{CalcHEP}. To avoid collinear and infrared divergences, we limit the phase space in the event generation with the following cuts on the outgoing photon energy $E_\gamma$ and its polar angle $\theta_\gamma$: 
\begin{equation} \label{eq:CutsPh}
 8\text{ GeV} < E_{\gamma} < 500\text{ GeV}, \;\;\;\; |\cos\theta_\gamma|\le 0.995 \, .
\end{equation}
The \emph{radiative} Bhabha scattering events are generated using \texttt{WHIZARD}~\cite{Kilian:2007gr} (to better handle the singularities)  with the same set of cuts as in Eq.~\eqref{eq:CutsPh} to the matrix element photon (i.e., excluding the ISR and  beamsstrahlung photons). Also, some additional cuts are implemented for the Bhabha process to take care of the soft and collinear divergences: 
\begin{align}
M_{e^{\pm}_{\rm in}, e^{\pm}_{\rm out}}<2m_e,~M_{e^{\pm}_{\rm out}, e^{\pm}_{\rm out}}<5\text{ GeV},~P_T^{\gamma}>1\text{ GeV},~\Delta R_{e^{\pm},\gamma}>0.2,~\Delta R_{e^{\pm},e^{\pm}}>0.4.
\end{align}

\begin{figure}[!t]
\centering
\includegraphics[width=0.49\textwidth]{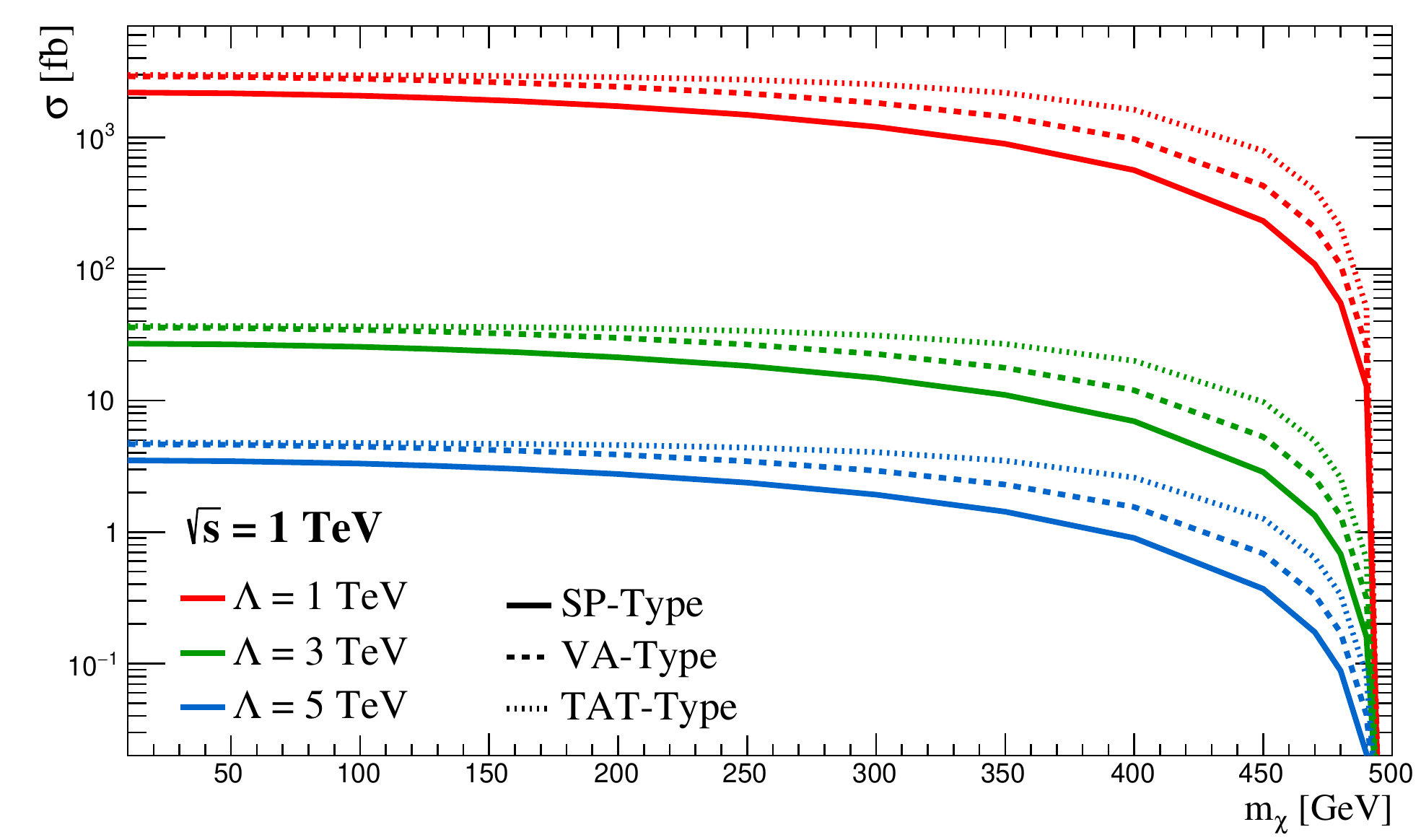}
\includegraphics[width=0.49\textwidth]{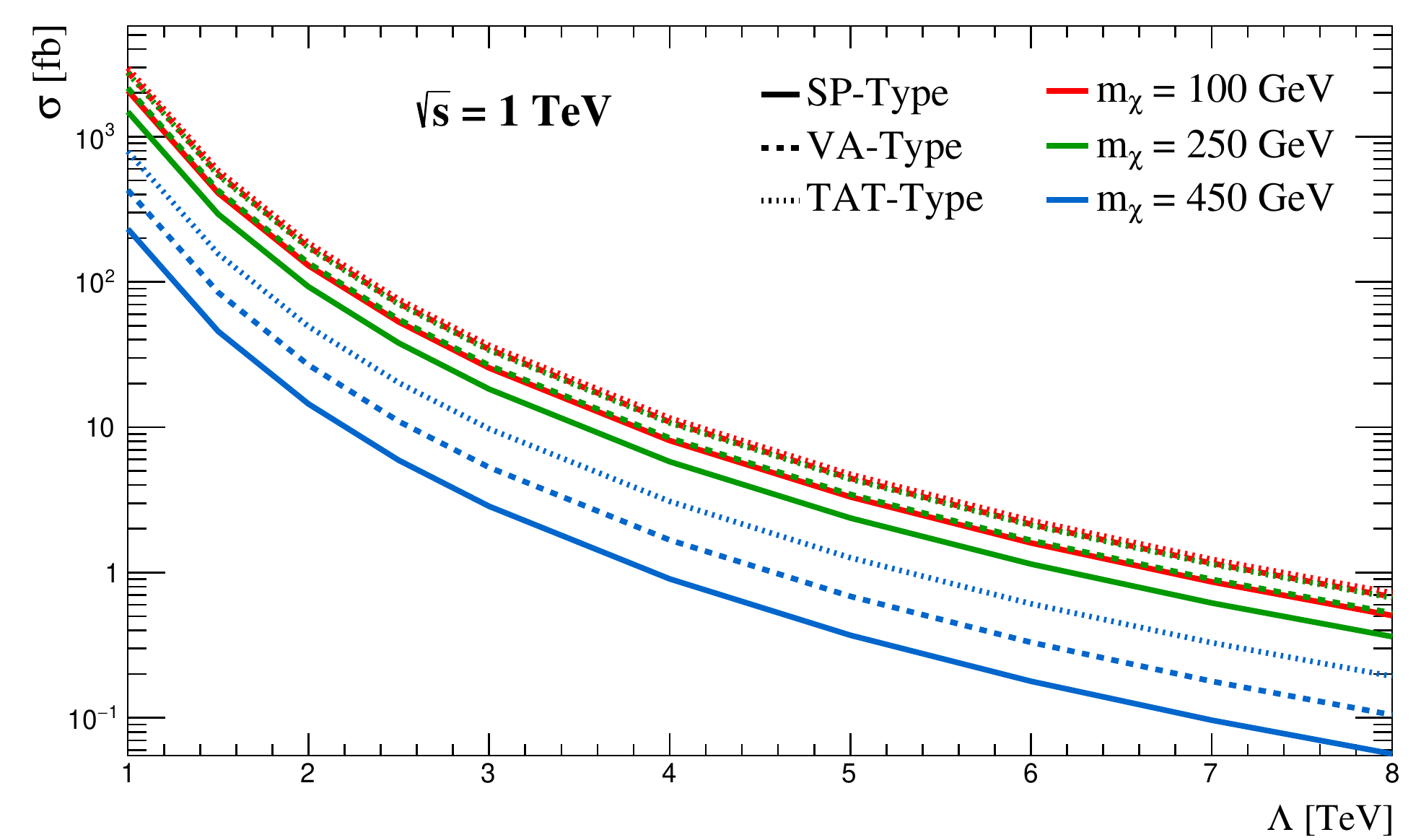}
\caption{Variation of mono-photon signal cross-section with the DM mass (left) and the cut-off scale (right) at $\sqrt s=1$ TeV ILC. The solid, dashed and dotted lines are for the SP, VA and TAT-type operators respectively. In the left panel, the red, green and blue curves respectively correspond to different values of the cut-off scale $\Lambda=1$ TeV, 3 TeV and 5 TeV, while in the right panel, they correspond to different values of the DM mass $m_\chi=100$ GeV, 250 GeV and 450 GeV.}
\label{fig:gCS}
\end{figure}
After generating the signal and background events, we perform a fast detector simulation of the \texttt{SiD} detector of ILC~\cite{Behnke:2013lya} using \texttt{Delphes3}~\cite{deFavereau:2013fsa} with the configuration card validated in Ref.~\cite{Potter:2016pgp}. 
The variations of the unpolarized signal cross section as a function of the DM mass and the cut-off scale are shown in Fig.~\ref{fig:gCS} left and right panels respectively for all three operator types, namely, SP (solid), VA (dashed) and TAT (dotted)-type. We find that the cross-section is the smallest (largest) for the SP (TAT)-type operator at any given DM mass. In the left panel, the sudden drop in the cross-section as $m_\chi$ approaches $\sqrt s/2$ is due to phase-space suppression. Otherwise, for smaller DM masses, the cross-section for a given operator type and a given cut-off scale is almost independent of the DM mass. In the right panel, we see that for a given DM mass the cross-section drops as $\Lambda^{-4}$, as expected.      

As for the background, we find that the neutrino background cross section at $\sqrt s=1$ TeV is 4.8 pb, while the radiative Bhabha background is 68.4 pb (though it is substantially reduced after the baseline selection). On the other hand, the DM signal cross section is found to be much smaller, as shown in Table~\ref{table:PolG} for a benchmark DM mass of $m_\chi=100$ GeV and the cut-off scale $\Lambda=3$ TeV. 
\begin{table}[!t]
 \centering 
 \tbl{Mono-photon background and signal cross-sections with different beam polarizations at $\sqrt{s} = 1~\rm{TeV}$. For the signal, we have fixed $m_\chi = 100~\rm{GeV}$ and $\Lambda = 3~\rm{TeV}$. The numbers in bold highlight the optimal polarization choice for a given operator type.}
{ \begin{tabular}{|c|c|c|cccc|}
\hline
 \textbf{Process} & \textbf{Unpolarized} & \textbf{Polarization} &  \multicolumn{4}{c|}{\textbf{Polarized cross-section (fb)}}   \\
   \cline{4-7}
 \textbf{type} & \textbf{cross-section (fb)} & \textbf{\boldmath{$P(e^{-},e^{+})$}} & $(+,+)$ & $(+,-)$ & $(-,+)$ & $(-,-)$ \\
\hline
\hline
           &  & $(80,0)$ & 1106 & 1106 & 8506 & 8506 \\
$\nu\overline{\nu}\gamma$ & $4782$ & $(80,20)$ & 1268 & 963 & 10160  & 6793   \\
           &  & $(80,30)$ & 1393 & 860 & 10993 & 5931 \\

\hline
           &  & $(80,0)$ & 67920 & 67920 & 68867 & 68867    \\           
$e^- e^+ \gamma$ & $68439$ & $(80,20)$ & 67909 & 68386 & 69285  & 68297   \\
           &  & $(80,30)$ & 67809 & 68566 & 69502 & 68181 \\

\hline

           &  & $(80,0)$ & 25.5 & 25.5 & 25.5 & 25.5 \\
SP-type & $25.5$ & $(80,20)$ & 29.6 & 21.4 & 21.4 & 29.6 \\
           &  & $(80,30)$ & \bf{31.6} & 19.4 & 19.4 & 31.6 \\

\hline

           &  & $(80,0)$ & 61.7 & 61.7 & 6.9 & 6.9 \\
VA-type & $34.3$ & $(80,20)$ & 49.4 & 74.1 & 5.5 & 8.2 \\
           &  & $(80,30)$ & 43.2 & \bf{80.3} & 4.8 & 8.9 \\

\hline

           &  & $(80,0)$ & 36.5 & 36.5 & 36.5 & 36.5 \\
TAT-type & $36.5$ & $(80,20)$ & 42.3 & 30.6 & 30.6 & 42.3 \\
           &  & $(80,30)$ & \bf{45.2} & 27.7 & 27.7 & 45.2 \\

\hline
\end{tabular}}
 \label{table:PolG}
\end{table}  
%
\subsection{Effect of polarization} \label{sec:3.2} 

One important advantage of lepton colliders is that the incoming beams can be polarized. This helps to reduce the neutrino background considerably, as shown in Table~\ref{table:PolG}. To utilize the full advantage of the beam polarization, we investigate the effect of different choices of polarization on the signal and background. At the ILC, the baseline design foresees at least 80\% electron beam polarization at the interaction point, whereas the positron beam can be polarized up to 30\% for the undulator positron source (up to 60\% may be possible with the addition of a photon collimator)~\cite{Bambade:2019fyw}. 
For comparison, we show our results for three different nominal absolute values of polarization: $|P(e^-,e^+)|=(80,0)$,  (80,20) and (80,30). In each case, we can also have four different polarization configurations, namely, ${\rm sign}(P(e^-),P(e^+))=(+,+)$, $(+,-)$, $(-,+)$ and $(-,-)$, where $+$ and $-$ denote the right- and left-handed helicities respectively. 

In Table~\ref{table:PolG}, we show the effect of different schemes of polarizations and helicity orientations on the mono-photon signal and background cross-sections. It is clear that the radiative Bhabha background remains almost unchanged. On the other hand, electron beam polarization is very effective in reducing the neutrino background, as a 80\% \emph{right-handed} electron beam can reduce the neutrino background to $23\%$ of the unpolarized case, even without any polarization on the positron beam. The effect is further enhanced by a \emph{left-handed} positron beam. We see that for $20\%$ and $30\%$ left-handed positron beam polarization, the neutrino background is reduced to $20\%$ and $18\%$ of its unpolarized value, respectively. 

The signals are also affected to some extent by beam polarization and the optimal helicity configuration depends on the operator type. For SP- and TAT-type operators we see no effect of electron-beam polarization, but a 20\% (30\%) \emph{right-handed} positron beam can enhance the signal by 16\% (24\%). The VA-type signal, on the other hand, prefers the $(+,-)$ helicity configuration -- the same choice for which the neutrino background is minimized. With the $(+80\%,-30\%)$ configuration, the VA-type signal is enhanced by a factor of 2.3, whereas the $(+80\%,+30\%)$ configuration enhances it by a modest 26\%. 

Overall, although the $(+80\%,-30\%)$ configuration minimizes the background the most, looking at the different signal to background ratio, we find that the $(+80\%,+30\%)$ configuration is the best for the SP- and TAT-type operators. For direct comparison between the results for different operators, we choose to work with the $(+80\%,+30\%)$ configuration democratically for all the operator types, unless otherwise specified. 
%
%
\subsection{Cut-based analysis}\label{sec:3.3}
Now we analyze various kinematic distributions and perform a cut-based analysis to optimize the signal-to-background ratio. This of course depends on the DM mass, so in Table~\ref{table:BPs&CutsGamma}, we list three benchmark points (BPs) with $m_\chi=100$ GeV, 250 GeV and 350 GeV respectively, and present the corresponding selection cuts optimized for each case. Here we fix $\Lambda=3$ TeV for illustration, but in the next subsection, we will vary both $m_\chi$ and $\Lambda$ to obtain the $3\sigma$ sensitivity limits. As for the choice of the DM mass values, since it was seen from Figure \ref{fig:gCS} that the signal cross-sections are barely sensitive to the DM mass up to around $100$ GeV, our BP1 essentially captures the light DM scenario. Similarly, our BP3 is chosen moderately close to the kinematic limit of $\sqrt s/2$ (going too close to $\sqrt s/2$ will result in cross-section values too low too low to give sizable event counts after all the selection cuts). The BP2 is chosen for an intermediate mass DM in between BP1 and BP2. 
%
\begin{table}[t]
 \centering
 \small
 \tbl{Mono-photon selection cuts for different BPs across all operator types. }
{ \begin{tabular}{|c|c|c|c|}
  \hline
  & {\bf BP1} & {\bf BP2} & {\bf BP3} \\
  \hline\hline 
   &&& \\[-1.2em]
  \multirow{2}{4.5em}{\centering {\bf Definition}} & $m_{\chi} = 100\text{ GeV, }$ & $m_{\chi} = 250\text{ GeV, }$ & $m_{\chi} = 350\text{ GeV, }$ \\
  & $\Lambda = 3\text{ TeV}$ & $\Lambda = 3\text{ TeV}$ & $\Lambda = 3\text{ TeV}$ \\
  \hline
   &\multicolumn{3}{|c|}{}\\[-1.1em]
  Baseline selection & \multicolumn{3}{|c|}{$E_{\gamma} > 10\text{ GeV},\;\; |\eta_{\gamma}| < 2.45,\;\; P_T^{\rm miss}>10\text{ GeV}$} \\
  \hline\hline 
  \multicolumn{4}{|l|}{\textbf{SP-type}} \\
  \hline
  Cut-1 & $E_{\gamma} < 450\text{ GeV}$ & $E_{\gamma} < 340\text{ GeV}$ & $E_{\gamma} < 250\text{ GeV}$ \\
  \hline
  Cut-2 & \multicolumn{3}{|c|}{$|\eta_{\gamma}| < 1.6$} \\
  \hline
  Cut-3 & $ P_T^{\rm miss} < 450\text{ GeV} $ & $ P_T^{\rm miss} < 340\text{ GeV} $ & $ P_T^{\rm miss} < 240\text{ GeV} $ \\
  \hline
  Cut-4 &  \multicolumn{3}{|c|}{$ P_T^{\rm frac} < 1.3 $} \\
  \hline
   Cut-5 & \multicolumn{3}{|c|}{$1.1<\Delta R_{\gamma,{\rm MET}} < 4.5$} \\
  \hline

  \hline\hline 

\multicolumn{4}{|l|}{\textbf{VA-type}} \\
  \hline
  Cut-1 & $E_{\gamma} < 440\text{ GeV}$ & $E_{\gamma} < 350\text{ GeV}$ & $E_{\gamma} < 250\text{ GeV}$ \\
  \hline
  Cut-2 & \multicolumn{3}{|c|}{$|\eta_{\gamma}| < 1.7$} \\
  \hline
  Cut-3 & $ P_T^{\rm miss} < 400\text{ GeV} $ & $ P_T^{\rm miss} < 340\text{ GeV} $ & $ P_T^{\rm miss} < 250\text{ GeV} $ \\
  \hline
  Cut-4 &  \multicolumn{3}{|c|}{$ P_T^{\rm frac} < 1.2$} \\
  \hline
   Cut-5 & \multicolumn{3}{|c|}{$1.1<\Delta R_{\gamma,{\rm MET}} < 4.5$} \\

  \hline\hline 

  \multicolumn{4}{|l|}{\textbf{TAT-type}} \\
  \hline
  Cut-1 & $E_{\gamma} < 460\text{ GeV}$ & $E_{\gamma} < 360\text{ GeV}$ & $E_{\gamma} < 230\text{ GeV}$ \\
  \hline
  Cut-2 & \multicolumn{3}{|c|}{$|\eta_{\gamma}| < 1.7$} \\
  \hline
  Cut-3 & $ P_T^{\rm miss} < 450\text{ GeV} $ & $ P_T^{\rm miss} < 350\text{ GeV} $ & $ P_T^{\rm miss} < 230\text{ GeV} $ \\
  \hline
  Cut-4 &  \multicolumn{3}{|c|}{$ P_T^{\rm frac} < 1.2$} \\
  \hline
   Cut-5 & \multicolumn{3}{|c|}{$1.1<\Delta R_{\gamma,{\rm MET}} < 4.4$} \\
   \hline 
 \end{tabular} }

\label{table:BPs&CutsGamma}
\end{table}

%
We define our mono-photon signals by those events that pass through the baseline selection criteria as defined below, in addition to the cuts given in Eq.~\eqref{eq:CutsPh}:
\begin{equation}
E_{\gamma} > 10\text{ GeV},\;\; |\eta_{\gamma}| < 2.45\;\text{ and }\; P_T^{\rm miss}>10\text{ GeV} \, ,
\label{eq:baseline}
\end{equation}
where the hardest photon in an event is considered as the signal photon. For the radiative Bhabha  background, we define the selection criteria for electrons as $P_{T,e}>10\text{ GeV, }|\eta_e|<2.5$, and have kept only those events which contain no electrons (and positrons) passing these criteria, which means they have escaped detection. After implementing these baseline selection cuts, we find that the signal and the neutrino background are reduced to about 60\% of their original values in Table~\ref{table:PolG}. Similarly, the actual Bhabha-induced background relevant for our signal is found to be only about 13\% of its original value quoted in Table~\ref{table:PolG} after the baseline selection cuts, taking into account only the missed electron events. To further enhance our signal-to-background ratio, we then examine the signal versus background distributions of some relevant kinematic variables and devise further cuts,  which are dynamic with respect to different BPs, as summarized in Table \ref{table:BPs&CutsGamma}. See Ref.~\cite{Kundu:2021cmo} for details.

Even after implementing the baseline and analysis cuts 1 through 5, the neutrino background can only be reduced to about 40\% of its original value in Table~\ref{table:PolG}. Similarly, the radiative Bhabha background, although substantially reduced to about 4\% of its original value in Table~\ref{table:PolG} after the baseline selection and analysis cuts, still remains sizable and comparable to the neutrino background.  However, an electromagnetic calorimeter in the very forward direction of the beamline (BeamCal)~\cite{Abramowicz:2010bg} can further suppress the Bhabha background to the per mille level. To properly incorporate the effect of BeamCal, we have used the selection efficiencies obtained from a full detector simulation performed in Ref.~\cite{Habermehl:2020njb} by modeling the complete instrumented region in a realistic way. According to this analysis, the selection efficiency of the Bhabha background after the BeamCal veto only is $2.7\%$, while that of the neutrino background is between 98\% and 99.6\%. As for the DM signal, we expect it to be basically unaffected (just like the neutrino background) by the BealCal veto, as it does not contain highly energetic charged particles in the longitudinal direction.

For the polarized case, after the baseline selection cuts, the Bhabha background remains almost same as in the unpolarized case. The neutrino background, on the other hand, is significantly reduced in the polarized case to about 28\% of its unpolarized value. The other cut efficiencies are also slightly better for the neutrino background in the polarized case.   

As for the signals, from Table~\ref{table:PolG}, we see that the TAT-type operator has the largest cross section to start with, both for the unpolarized as well as for the $(+80\%, +30\%)$ polarized cases. Even after the baseline selection and the specialized cuts discussed above, the TAT-type signal retains the largest efficiency among the three types. This will be reflected in our signal significance results below.

\subsection{Signal significance} \label{sec:3.4}
After implementing all the cuts mentioned above, we calculate the final signal significance for our benchmark scenarios using the definition
\begin{equation}
    {\rm Sig} = \frac{S}{\sqrt{S+B+(\epsilon B)^2}} \, ,
     \label{eq:significance}
\end{equation}
where $S$ and $B$ are the number of signal and total background events respectively for a given integrated luminosity, and  $\epsilon$ is the background systematic uncertainty. Our results are given in Table~\ref{table:SigG} for the three BPs. We show the numbers for an ideal case with zero systematics and also for a more realistic case with 1\% systematics, i.e. with  $\epsilon=0.01$ (in parentheses). The results are significantly weakened in the latter case because of the relatively large background compared to the signal.   

\begin{table}[t!]
  \centering
   \tbl{Signal significance  in the mono-photon 
  channel for the three BPs at $\sqrt{s}=1$ TeV. The values in parenthesis correspond to 1\% background systematic  uncertainty.}
{
  \begin{tabular}{|l|ccc||ccc|}
  \hline
  \multirow{3}{3.5em}{\bf Operator type} & \multicolumn{6}{|c|}{\bf Signal significance for ${\cal L}_{\rm int}=1000\,{\rm fb}^{-1}$}\\
   \cline{2-7}
   & \multicolumn{3}{|c||}{Unpolarized beams} & \multicolumn{3}{|c|}{Polarized beams}\\
   \cline{2-7}
   & BP-1 & BP-2 & BP-3 & BP-1 & BP-2 & BP-3  \\
  \hline
  \hline
  SP-type & $8.1\;(0.6)$ & $5.8\;(0.4)$ & $3.5\;(0.3)$ 
  & $18.1\;(2.4)$ & $13.0\;(1.7)$ & $7.8\;(1.0)$  \\
  \hline
  VA-type & $10.9\;(0.8)$ & $8.5\;(0.6)$ & $5.6\;(0.4)$ 
  & $24.9\;(3.2)$ & $19.4\;(2.5)$ & $12.9\;(1.7)$  \\
  \hline
  TAT-type & $11.8\;(0.8)$ & $10.8\;(0.8)$ & $8.5\;(0.6)$ 
  & $26.2\;(3.5)$ & $24.1\;(3.2)$ & $19.2\;(2.6)$  \\
  \hline 
 \end{tabular}
 }
  \label{table:SigG}
\end{table}

From Table~\ref{table:SigG}, we see that the significance enhances as we go to lower DM mass regions, as expected because of kinematic reasons. Operator-wise we see that TAT and VA-type operators perform better than the SP-type. We also find substantial (around 50\%) increase in significance on application of optimal beam polarization. 

\begin{figure}[t]
\centering 
\includegraphics[width=0.35\linewidth]{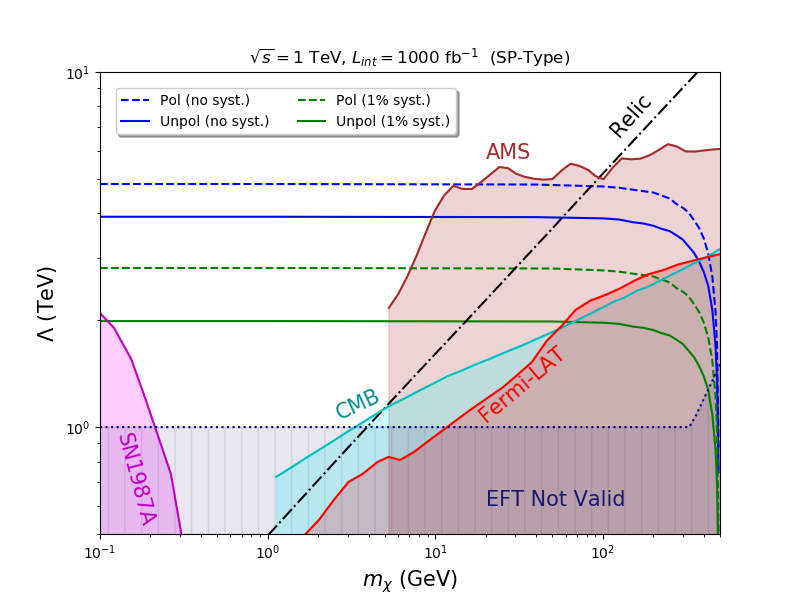}
\hspace{-0.7cm}
\includegraphics[width=0.35\linewidth]{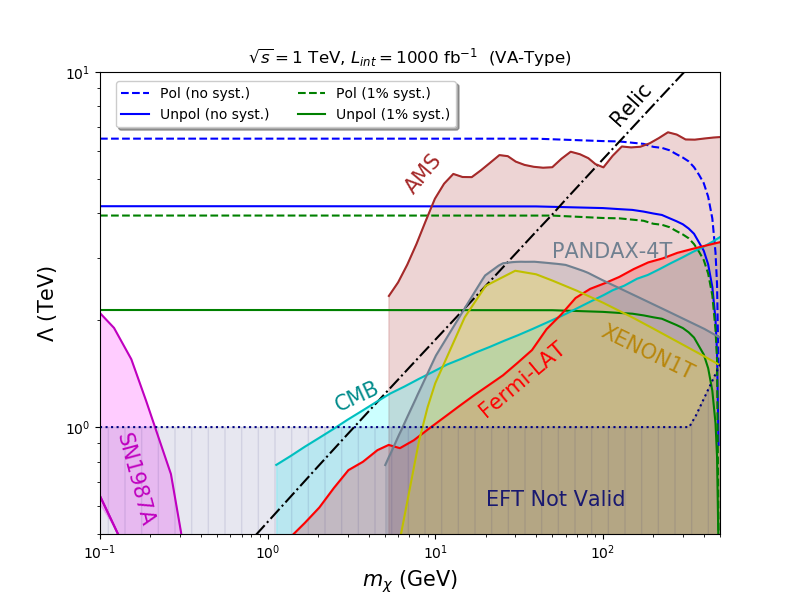}
\hspace{-0.7cm}
\includegraphics[width=0.35\linewidth]{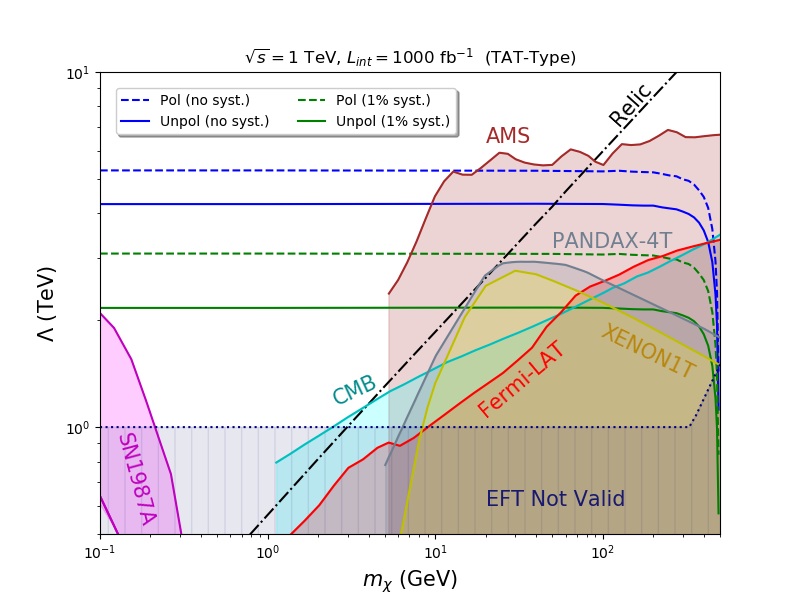}
\caption{$3\sigma$ sensitivity contours in the mono-photon channel for the SP (left), VA (middle) and TAT (right)-type operators with \emph{unpolarized} (solid lines) and \emph{polarized} (dashed lines) $e^+e^-$ beams at $\sqrt s=1$ TeV center-of-mass energy and with  ${\cal L}_{\rm int}=1000$ fb$^{-1}$ integrated luminosity. The blue (green) contours are assuming zero (1\%) background systematics. The various shaded regions are excluded by direct detection (XENON1T, PANDAX-4T), indirect detection (Fermi-LAT, AMS), astrophysics (SN1987A) and cosmology (CMB) constraints. In the shaded region below $\Lambda={\rm max}\{\sqrt s/2,3m_\chi\}$, our EFT framework is not valid. Along the dot-dashed line, the observed DM relic density is reproduced for a thermal DM assuming only DM-electron effective coupling. }
\label{figure:ContourG}
\end{figure}

Going beyond the three BPs, we now vary the DM mass and calculate the signal significance following the same cut-based analysis procedure outlined above. Our results for the $3\sigma$ sensitivity contours in the $(m_\chi,\Lambda)$ plane are shown in Figure \ref{figure:ContourG} for all the operator types. The solid (dashed) contours are for the unpolarized (optimally polarized) case, and the blue (green) contours are assuming zero (1\%) background systematics. The shaded regions are excluded by various constraints. First of all, for $\Lambda<{\rm max}\{\sqrt s/2,3m_\chi\}$, our EFT framework is not valid (cf.~Sec.~\ref{sec:2}). This is shown by the navy blue-shaded regions in Fig.~\ref{figure:ContourG}. For $\sqrt s=1$ TeV as considered here, this EFT validity limit supersedes the previous LEP limit~\cite{Fox:2011fx}.  

The same effective operator given in Eq.~\eqref{eq:EFT} also gives rise to DM scattering with electrons $\chi e^-\to \chi e^-$. The exact analytic expressions for these cross sections in our EFT framework can be found in Appendix C of Ref.~\cite{Guha:2018mli} for all the operator types. 
Comparing these with the experimental upper limits on $\sigma_{\chi e}$ from dedicated direct detection experiments~\cite{XENON100:2015tol, XENON:2019gfn}, we can derive a {\it lower} limit on the cut-off scale $\Lambda$ as a function of the DM mass $m_\chi$. However, the current best limit on $\sigma_{\chi e}$ from XENON1T is at the level of ${\cal O}(10^{-39})\:{\rm cm}^2$~\cite{XENON:2019gfn}, which translates into a very weak bound on $\Lambda$ and is not relevant for our study. Even the future ambitious proposals like DARKSPHERE can only reach up to ${\cal O}(10^{-42})\:{\rm cm}^2$~\cite{Hamaide:2021hlp}, still 5 orders of magnitude weaker than that needed to probe a TeV-scale $\Lambda$ value. 

However, more stringent limits can be derived from DM-nucleon scattering searches. Even for an LDM as in our case, DM-nucleon couplings are necessarily induced at loop level from photon exchange between virtual leptons and the quarks. In fact, as shown in Ref.~\cite{Kopp:2009et}, the loop-induced DM-nucleon scattering almost always dominates over the DM-electron scattering. The analytic expressions for the one and two-loop DM-nucleon scattering cross sections can be found in Ref.~\cite{Kopp:2009et}. 
We have translated the experimental upper limits from XENON1T~\cite{XENON:2018voc} and PANDAX-4T~\cite{PandaX-4T:2021bab} onto the $(m_\chi,\Lambda)$ plane, as shown by the yellow and grey-shaded regions respectively in Fig.~\ref{figure:ContourG}. Note that these limits are only applicable for the vector and tensor lepton currents, i.e. $\Gamma_\ell=\gamma_\mu,\ \sigma_{\mu\nu}$ in Eq.~\eqref{eq:EFT}. For the scalar lepton current, $\Gamma_\ell=1$, the one-loop DM-nucleon coupling vanishes, and one has to go to two loops which is suppressed by $\alpha_{\rm em}^2$ for the S-S type coupling and $\alpha_{\rm em}^2v^2$ (where $v\sim 10^{-3}$ is the DM velocity) for the P-S type coupling. In contrast, for pseudo-scalar and axial-vector lepton currents, i.e. $\Gamma_\ell=\gamma_5, \ \gamma_\mu \gamma_5$, the DM-nucleon coupling vanishes to all orders. Therefore, we have not shown the XENON1T and PANDAX-4T limits for the SP-type operator on the top left panel of Fig.~\ref{figure:ContourG}.

The same effective operator given in Eq.~\eqref{eq:EFT} also enables DM annihilation into electrons $\chi \overline{\chi} \to e^+ e^-$. The exact analytic expressions for these cross sections in our EFT framework can be found in Appendix C of Ref.~\cite{Guha:2018mli} for all the operator types. Using these, we calculate the thermal-averaged cross section times relative velocity $\langle \sigma v\rangle$ which goes as $m_\chi^2/\Lambda^{4}$ and compare it with the existing indirect detection upper limits on $\langle \sigma v\rangle$ in the $e^+e^-$ channel to put a lower bound on $\Lambda$ as a function of the DM mass. This is shown in Fig.~\ref{figure:ContourG} by the red and brown-shaded regions respectively for the Fermi-LAT~\cite{Leane:2018kjk} and AMS-02~\cite{John:2021ugy} constraints on $\langle \sigma v\rangle$. Similar constraints on $\langle \sigma v\rangle$ can be derived using CMB anisotropies~\cite{Leane:2018kjk}, which is shown by the cyan-shaded region in Fig.~\ref{figure:ContourG}, assuming an $s$-wave annihilation (for $p$-wave annihilation, the CMB bound will be much weaker).  

Along the dot-dashed line in Fig.~\ref{figure:ContourG}, the observed relic density can be reproduced for a DM. In principle, the region to the left and above of this line is disfavored for a thermal DM, because in this region $\langle \sigma v\rangle$ is smaller than the observed value of $\sim (2-5)\times 10^{-26}{\rm cm}^3{\rm sec}^{-1}$ (depending on the DM mass~\cite{Steigman:2012nb}), which leads to an overabundance of DM, since $\Omega_\chi h^2 \propto 1/\langle \sigma v\rangle$. However, this problem can be circumvented by either opening up additional leptonic annihilation channels (like $\mu^+\mu^-$,  $\tau^+\tau^-$ and $\nu\bar{\nu}$) or even going beyond the DM paradigm and invoking e.g., the freeze-in mechanism~\cite{Hall:2009bx}. This will not affect our main results, since the collider phenomenology discussed here only depends on the DM coupling to electrons. 

Also shown in Fig.~\ref{figure:ContourG} is the supernova constraint, which excludes the magenta-shaded region from consideration of energy-loss and optical depth criteria from the observation of SN1987A~\cite{Guha:2018mli}. Here we have used an average supernova core temperature of 30 MeV. Note that the supernova bound is only applicable for DM mass below $\sim$ 200 MeV or so, and for a certain range of $\Lambda$ values, above which the DM particles cannot be efficiently produced in the supernova core, and below which they will no longer free-stream.

%
%

From Fig.~\ref{figure:ContourG}, we find that in spite of a large irreducible background, the accessible range of the cut-off scale $\Lambda$ at $\sqrt s=1$ TeV ILC looks quite promising in the mono-photon channel, especially for low mass DM, where the collider sensitivity is almost flat, whereas the existing direct and indirect detection constraints are much weaker. This complementarity makes the collider searches for DM very promising. With unpolarized beams, the 3$\sigma$-reach for the SP-type operator can be up to 3.9 TeV, while for the VA and TAT-type operators, it can be up to 4.2 TeV. With optimally polarized beams, i.e. with $(+80\%,+30\%)$ for the SP and TAT-types and $(+80\%,-30\%)$ for the VA type, the sensitivity reaches can be extended to $4.8$ TeV (SP), $6.5$ TeV (VA) and $5.3$ TeV (TAT), as shown in Fig.~\ref{figure:ContourG}.



%

%
%
\section{Mono-$Z$ channel} \label{sec:4}
In addition to the mono-photon channel discussed in the previous section, another useful channel for LDM search at lepton colliders is the mono-$Z$ channel, where the $Z$-boson is emitted from one of the initial states. Depending on the subsequent decay of the $Z$-boson to either leptonic or hadronic final states, we perform a dedicated cut-based signal and background analysis, as discussed below.     

\subsection{Leptonic mode} \label{sec:4.1}
For the leptonic decay of the $Z$-boson, we examine the process $e^+e^-\to\chi\overline{\chi}Z(\to \ell^-\ell^+)$. We will only consider $\ell=e,\mu$ for simplicity and use the lepton pair as the visible particles for tagging. The main SM background for this channel is  $e^+e^-\to\nu\overline{\nu}\;\ell^+\ell^-$, and it is polarization-dependent. 
\subsubsection{Unpolarized and polarized cross-sections}
For the signal and background simulation, we generated the UFO library for our EFT framework using \texttt{FeynRules}~\cite{Alloul:2013bka} and then generated events for both signal and background using \texttt{MadGraph~5}~\cite{Alwall:2014hca} with the following basic baseline cuts:
\begin{equation}
P_T(\ell) > 10\text{ GeV},\;\;\;\; |\eta_\ell| \le 2.5,\;\;\;\; \Delta R_{\ell\ell} \ge 0.4  \, .
\end{equation}
For the signal, the $Z$-bosons are decayed into the charged lepton pairs via the \texttt{MadSpin}~\cite{Frixione:2007zp, Artoisenet:2012st} package which is implemented in \texttt{MadGraph 5}, to take care of the spin-correlation effects of the lepton pairs. A fast detector simulation to these events is done using \texttt{Delphes~3}~\cite{deFavereau:2013fsa} with the same configuration card~\cite{Potter:2016pgp} as in Sec.~\ref{sec:3.1}. 

With unpolarized beams, we find that the neutrino background cross section at $\sqrt s=1$ TeV is 420.5 fb, whereas the DM signal cross section is much smaller, as shown in Table~\ref{table:PolZ} for a benchmark DM mass of $m_\chi=100$ GeV and the cut-off scale $\Lambda=3$ TeV.  Similar to the mono-photon case, we also examine the effect of polarization on the signal and background cross-sections, as shown in Table~\ref{table:PolZ}. The neutrino background can be reduced to 28\% of its original value by making the electron beam $+80\%$ polarized, and further reduced to 21\% of its original value by additionally making the positron beam $-30\%$ polarized. The $(+80\%, -30\%)$ polarization configuration also enhances the VA-type signal by a factor of 2.4. However, the $(+80\%, +30\%)$ configuration is better for the SP and TAT-type signals. For ease of comparison between different operator types, we choose to work with the $(+80\%, +30\%)$ configuration democratically for all operator types, as well as for the background, unless otherwise specified.
\begin{table}[t!]
 \centering
\tbl{Comparison of the leptonic mono-$Z$  background and signal cross-sections for different choices of beam polarization for $m_{\chi}=100\text{ GeV}$ and $\Lambda=3\text{ TeV}$ at $\sqrt{s}=1\text{ TeV}$ ILC. The numbers in bold highlight the optimal polarization choice for a given operator type.}
{%
  \begin{tabular}{|c|c|c|cccc|}
  \hline
  \textbf{Process} &  \textbf{Unpolarized} & \textbf{Polarization} & \multicolumn{4}{|c|}{\textbf{Polarized cross-section (fb)}} \\ \cline{4-7} 
 \textbf{type} & \textbf{cross-section (fb)} & \textbf{\boldmath{$P(e^-,e^+)$}} & \textbf{$(+,+)$} & \textbf{$(+,-)$} & \textbf{$(-,+)$} & \textbf{$(-,-)$} \\
  \hline
  \hline
  \multirow{3}{3.2em}{\centering $\nu\overline{\nu}\ell^-\ell^+$} &  & $(80, 0)$ & $116$ & $116$ & $723$ & $723$  \\
   &$420$ & $(80,20)$ & $135$ & $98$ & $856$ & $590$  \\
   & & $(80,30)$ & $145$ & $88$ & $926$ & $523$ \\
  \hline 
  \multirow{3}{5.2em}{\centering SP-Type} &  & (80, 0) & $0.26$ & $0.26$ & $0.25$ & $0.25$  \\
   &$0.28$ & (80,20) & $0.29$ & $0.22$ & $0.21$ & $0.29$  \\
   &  & $(80,30)$ & \boldmath{$0.32$} & $0.19$ & $0.19$ & $0.32$  \\
  \hline
  \multirow{3}{5.2em}{\centering VA-Type} &  & (80, 0) & $0.15$ & $0.15$ & $0.02$ & $0.02$  \\
   &$0.08$ & (80,20) & $0.12$ & $0.18$ & $0.01$ & $0.02$  \\
   &  & $(80,30)$ & $0.11$ & \boldmath{$0.19$} & $0.01$ & $0.02$  \\
  \hline 
  \multirow{3}{5.2em}{\centering TAT-Type} &  & (80, 0) & $0.62$ & $0.62$ & $0.62$ & $0.62$  \\
   &$0.68$ & (80,20) & $0.72$ & $0.52$ & $0.52$ & $0.72$  \\
   &  & $(80,30)$ & \boldmath{$0.77$} & $0.47$ & $0.47$ & $0.77$  \\
  \hline 
  \end{tabular}
 }
 \label{table:PolZ}
\end{table}

%
%
%
\subsubsection{Cut-based analysis}
We define our signals by those events that pass through the baseline selection criteria as defined below: 
 $P_{T, \ell} > 20\text{ GeV, }\; |\eta_\ell| < 2.45$,  
 where the $Z$-boson is reconstructed by the condition that all final state lepton-pairs are oppositely charged and of same flavor (OSSF). 
Other selection criteria are dynamic with respect to different BPs. We have taken the same three BPs as in the mono-photon case to probe  different regions of the parameter space, namely, BP1 essentially represents all light DM region, BP3 represents the region close to the kinematic limit of $\sqrt s/2-m_Z$, whereas BP2 captures the intermediate DM mass region. 

After implementing the baseline selection cuts, we find that the background is reduced to about 40\% of its original value in Table~\ref{table:PolZ} for the unpolarized (polarized) case, whereas the signals are reduced to about 60\%-70\% of their original values.  We then consider various kinematic distributions for the signal and background, and devise some specialized selection cuts~\cite{Kundu:2021cmo}. We find that after applying all these cuts, we can still retain about 35\%-45\% of the signal, whereas the background is reduced to below percent level of the original values given in Table~\ref{table:PolZ}.


\subsubsection{Results}
After implementing all these cuts, we calculate the final signal significance for the three BPs using Eq.~\eqref{eq:significance}. Our results are given in Table~\ref{table:SigZ} for an integrated luminosity of ${\cal L}_{\rm int}=1000$ fb$^{-1}$. We see that as we go higher up in the DM mass the signal significance drops. We also find that the best-performing operator type is the TAT-type, for which more than $97\%$ of the background events are removed after all the selection cuts. For the signal we retain $58\%-61\%$ of the events, although for BP3 only $48\%$ remains. The SP-type operator also gives good results, where we retain $50\%-66\%$ of the signal across BPs and polarization choices, while removing more than $96\%$ of the background events. Even for VA-type we retain more than $50\%$ of the signal events and are able to cut down the background event yields to $11\%$. We also notice the positive effect of the beam polarization by which we achieve an enhancement of signal significance by more than $2$ times compared to the ones with unpolarized beams. For VA-type though the significance can be further increased for the polarized beam case by choosing the left-handed positron beams as is evident from Table \ref{table:PolZ}. 

\begin{table}[t]
  \centering
\tbl{Signal significance in the mono-$Z$ leptonic channel at $\sqrt{s}=1$ TeV and $\mathcal{L}_{\rm int} = 1000\, {\rm fb}^{-1}$. The values in the parenthesis correspond to 1\% background systematic  uncertainty.}
{\resizebox{\textwidth}{!}{%
  \begin{tabular}{|l|ccc||ccc|}
  \hline
  \multirow{3}{4.1em}{\centering \bf Operator Type} & \multicolumn{6}{|c|}{\bf Signal significance for ${\cal L}_{\rm int}=1000\,{\rm fb}^{-1}$}\\
   \cline{2-7}
   & \multicolumn{3}{|c||}{Unpolarized beams} & \multicolumn{3}{|c|}{Polarized beams}\\
   \cline{2-7}
   & BP-1 & BP-2 & BP-3 & BP-1 & BP-2 & BP-3  \\
  \hline
  \hline
  SP-type & $1.7\;(1.3)$ & $0.7\;(0.5)$ & $0.1\;(0.1)$ & $3.8\;(3.6)$ & $1.7\;(1.5)$ & $0.3\;(0.3)$ \\
  \hline
  VA-type & $0.2\;(0.1)$ & $0.1\;(0.1)$ & $0.1\;(0.1)$ & $0.5\;(0.4)$ & $0.4\;(0.3)$ & $0.2\;(0.2)$ \\
  \hline
  TAT-type & $4.5\;(3.9)$ & $2.4\;(1.9)$ & $0.6\;(0.5)$ & $9.4\;(9.1)$ & $5.4\;(5.1)$ & $1.3\;(1.3)$ \\
  \hline 
 \end{tabular}
 }}
  \label{table:SigZ}
\end{table}
%
%

%
\begin{figure}[t!]
\centering 
\includegraphics[width=0.35\linewidth]{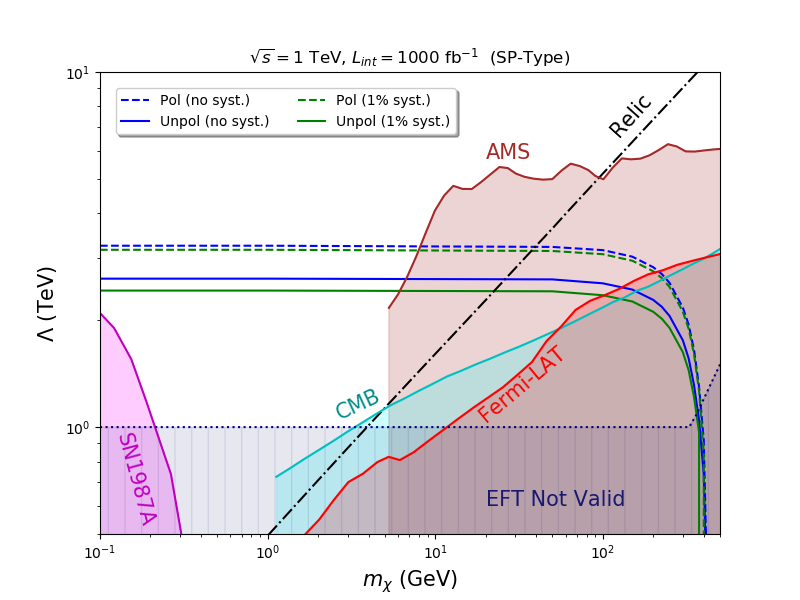}
\hspace{-0.7cm}
\includegraphics[width=0.35\linewidth]{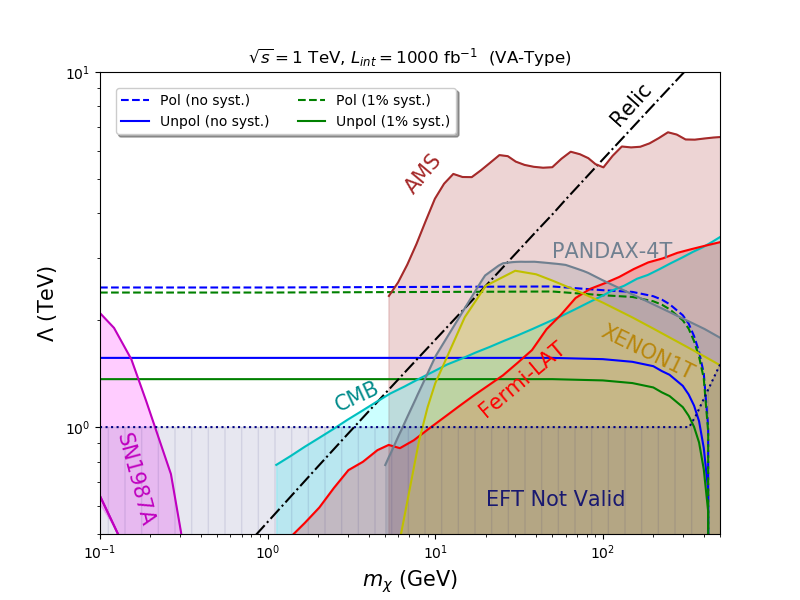}
\hspace{-0.7cm}
\includegraphics[width=0.35\linewidth]{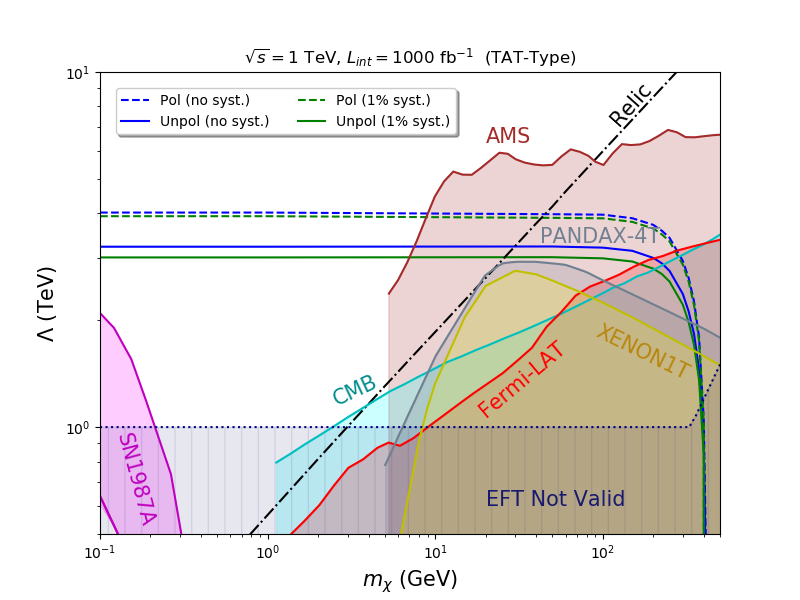}
\caption{$3\sigma$ sensitivity contours in the  mono-$Z$ leptonic channel. Labels are same as in Fig.~\ref{figure:ContourG}.}
\label{figure:ContourZ}
\end{figure}

Going beyond the three BPs, we now vary the DM mass and present the $3\sigma$ sensitivity reach for this channel in Fig.~\ref{figure:ContourZ} for all the operators. The labels and shaded regions are the same as in the mono-photon case (cf.~Fig.~\ref{figure:ContourG}). 
We see that the accessible range of the cut-off scale $\Lambda$ for the unpolarized beams can reach up to $3.2$~TeV for the TAT-type operator, whereas for the SP and VA-type, it can reach up to 2.6 TeV and 1.6 TeV respectively. But with the application of optimally polarized beams as discussed earlier, we see an increase by about $25\%$ of the $3\sigma$ reach on the $\Lambda$ scale, up to 3.2 TeV, 2.5 TeV and 4 TeV for the for SP, VA and TAT-type operators, respectively. 

\subsection{Hadronic mode} \label{sec:4.2}
Next we study  $e^+e^-\to\chi\overline{\chi}Z(\to jj)$, where $j\equiv u,d,c,s,b$ quarks. The relevant SM background processes for this channel are $e^+e^-\to\nu\overline{\nu}jj$ and $e^+e^-\to jj\ell \nu$ (with one charged lepton escaping the detector) where the jets and leptons in the final state can come from any possible source (not necessarily from an on-shell $Z$). %

\subsubsection{Unpolarized and polarized cross-sections}
We use the same UFO library as before which is implemented using \texttt{FeynRules}~\cite{Alloul:2013bka} and simulate the events for the signal and backgrounds via \texttt{MadGraph~5}~\cite{Alwall:2014hca} with the following basic cuts to the parameter space:
\begin{equation}
P_T(j,\ell) > 10\text{ GeV},\;\;\;\; |\eta_j| \le 3.0,\;\;|\eta_\ell| \le 2.5,\;\; \Delta R_{jj,\ell j} \ge 0.4  \, .
\end{equation}
For the signals, as in the leptonic case, the on-shell $Z$-bosons are decayed into the pairs of jets using the \texttt{MadSpin} package~\cite{Frixione:2007zp, Artoisenet:2012st}, implemented in \texttt{MadGraph 5}. Both the signal and background samples are hadronized using \texttt{Pythia8.2}~\cite{Sjostrand:2014zea} and then the final state jets are reconstructed with with \texttt{anti-$kT$}~\cite{Cacciari:2008gp} clustering algorithm with a minimum $P_T$ of $10$ GeV and a cone radius ($R)$ of 0.4 using the \texttt{FastJet}~\cite{Cacciari:2011ma} package. The fast detector simulation to these events are done using \texttt{Delphes~3}~\cite{deFavereau:2013fsa} with the same configuration card~\cite{Potter:2016pgp} as discussed in Sec.~\ref{sec:3.1}. 

With unpolarized beams, we find the neutrino-pair background is 798 fb, whereas the $jj\ell\nu$ background is 1186 fb. On the other hand, the DM signal is only at a few fb level, as shown in Table~\ref{table:PolZjj} for a benchmark DM mass of $m_\chi=100$ GeV and the cut-off scale $\Lambda=3$ TeV.    We then examine different choices of beam-polarization on both the event samples for this channel, as shown in Table~\ref{table:PolZjj}. We find that both backgrounds are polarization-dependent and fall off significantly for right-handed electron beam and with increasing  degree of polarization. 
We choose the polarization configuration $P(e^-,e^+)=(+80\%,+30\%)$ democratically over all the operator types. 
%

%

%
\begin{table}[t]
 \centering 
 \small
 \tbl{Hadronic mono-$Z$ background and signal cross-sections for different choices of beam polarization with $m_\chi = 100~\rm{GeV}$ and $\Lambda = 3~\rm{TeV}$ at $\sqrt{s} = 1~\rm{TeV}$ ILC. The numbers in bold highlight the optimal polarization choice for a given operator type.}
 {\begin{tabular}{|c|c|c|cccc|}
\hline
 \textbf{Process} & \textbf{Unpolarized} & \textbf{Pol.} &  \multicolumn{4}{c|}{\textbf{Polarized cross-section (fb)}}   \\
   \cline{4-7}
 \textbf{type} & \textbf{cross-section (fb)} & \textbf{\boldmath{$P(e^{-},e^{+})$}} & $(+,+)$ & $(+,-)$ & $(-,+)$ & $(-,-)$ \\
\hline
\hline
           &  & $(80,0)$ & 178 & 178 & 1415 & 1415    \\
$\nu\overline{\nu} j j$ & $798$ & $(80,20)$ & 206 & 151 & 1689  & 1134   \\
           &  & $(80,30)$ & 219  & 136 & 1833 & 989 \\

\hline
           &  & $(80,0)$ & 302 & 302 & 2061 & 2061    \\           
$j j \ell \nu$ & $1186$ & $(80,20)$ & 359 & 246 & 2446  & 1685   \\
           &  & $(80,30)$ & 386  & 216 & 2635 &  1492 \\

\hline

           &  & $(80,0)$ & 2.57 & 2.57 & 2.58 &  2.58  \\
SP-Type & $2.78$ & $(80,20)$ & 2.98 & 2.15 & 2.15 &  2.97  \\
           &  & $(80,30)$ & \bf{3.17} & 1.95 & 1.95 &  3.17 \\

\hline

           &  & $(80,0)$ & 1.35 & 1.35 & 0.15 &  0.15  \\
VA-Type & $0.83$ & $(80,20)$ & 1.07 & 1.61 & 0.12 &  0.18  \\
           &  & $(80,30)$ & 0.94 & \bf{1.76} & 0.10 &  0.19 \\

\hline

           &  & $(80,0)$ & 6.22 & 6.22 & 6.21 & 6.21   \\
TAT-Type & $6.77$ & $(80,20)$ & 7.23 & 5.23 & 5.24 & 7.21   \\
           &  & $(80,30)$ & \bf{7.72} & 4.73 & 4.73 &  7.69 \\

\hline
\end{tabular}}
 \label{table:PolZjj}
\end{table}  
\subsubsection{Cut-based analysis}
After obtaining the signal and background cross-sections as reported in Table~\ref{table:PolZjj}, we proceed with our cut-based analysis to optimize the signal significance. 
We select the events that contain at least two jets with the following transverse momentum and pseudorapidity requirements: 
$P_{T, j} > 20\text{ GeV, }\; |\eta_j| < 2.45$. 
The hardest two jets are required to reconstruct the $Z$-boson. Further selection cuts are applied some of which depend on the DM mass. So, as in the leptonic channel, we have taken the same three BPs with varying DM mass and impose dynamic cuts~\cite{Kundu:2021cmo}. 
%

\begin{table}[t]
  \centering
\tbl{Signal significances of the mono-$Z$ hadronic channel  at $\sqrt{s}=1~{\rm TeV}$ and $\mathcal{L}_{\rm int} = 1000~{\rm fb}^{-1}$. The values in parenthesis correspond to 1\% background systematic uncertainty.}
{\resizebox{\textwidth}{!}{%
  \begin{tabular}{|l|ccc||ccc|}
  \hline
  \multirow{3}{6em}{\centering \bf Operator types} & \multicolumn{6}{|c|}{\bf Signal significance for ${\cal L}_{\rm int}=1000\,{\rm fb}^{-1}$}\\
   \cline{2-7}
   & \multicolumn{3}{|c||}{Unpolarized beams} & \multicolumn{3}{|c|}{Polarized Beam}\\
   \cline{2-7}
   & BP-1 & BP-2 & BP-3 & BP-1 & BP-2 & BP-3  \\
  \hline
  \hline
  SP-type & $4.7\;(3.3)$ & $1.5\;(1.0)$ & $0.3\;(0.2)$ & $10.3\;(9.1)$ & $ 3.6\;(3.1)$ & $0.8\;(0.6)$  \\
  \hline
  VA-type & $0.4\;(0.2)$ & $0.3\;(0.1)$ & $0.1\;(0.1)$ & $0.2\;(0.1)$ & $0.1\;(0.1)$ & $0.1\;(0.04)$  \\
  \hline
  TAT-type & $14.2\;(10.4)$ & $5.8\;(3.7)$ & $1.2\;(0.7)$ & $27.7\;(25.6)$ & $12.9\;(11.1)$ & $2.8\;(2.3)$  \\
  \hline 
 \end{tabular}
}
}
  \label{table:SigZjj}
\end{table}
\begin{figure}[t!]
\centering 
\includegraphics[width=0.35\linewidth]{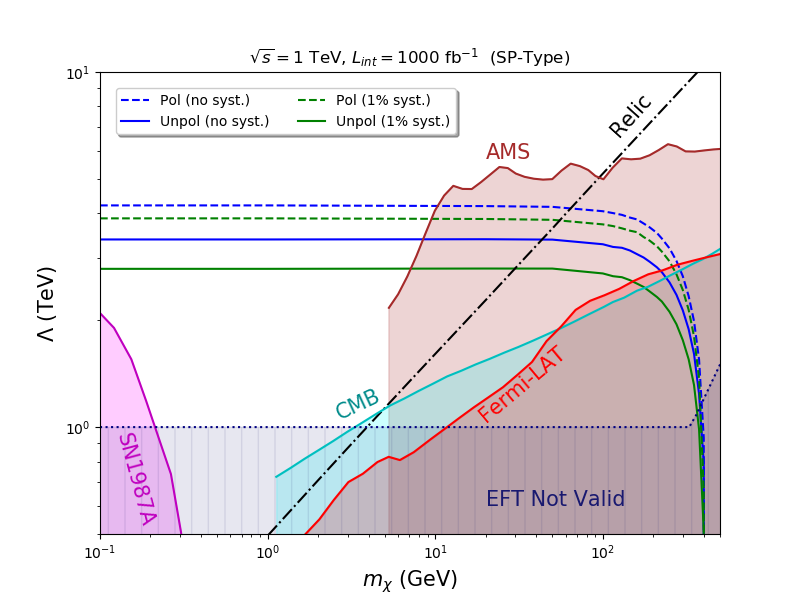}
\hspace{-0.7cm}
\includegraphics[width=0.35\linewidth]{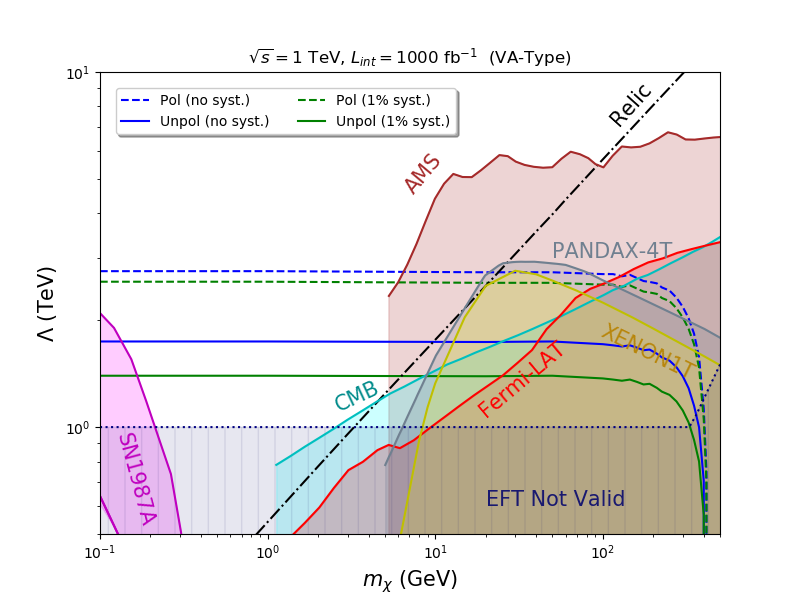} 
\hspace{-0.7cm}
\includegraphics[width=0.35\linewidth]{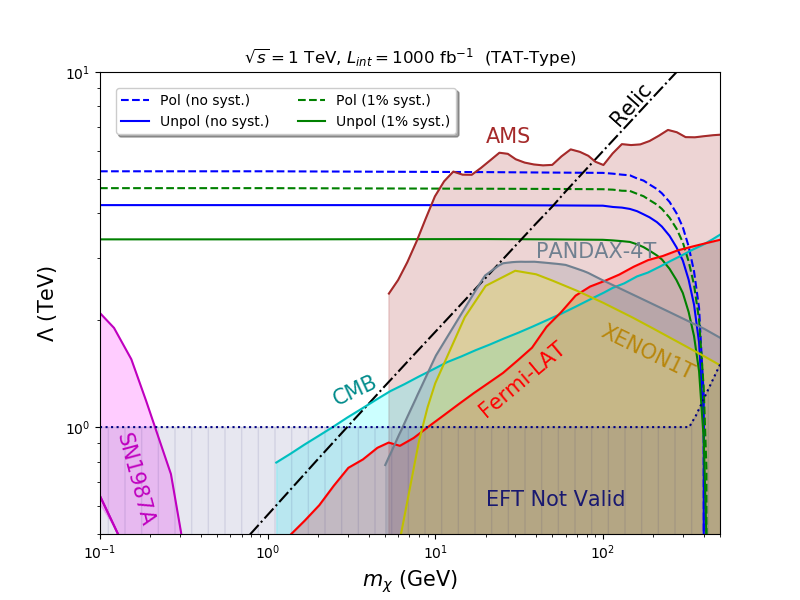}
\caption{$3\sigma$ sensitivity contours in the mono-$Z$ hadronic channel. Labels are same as in Fig.~\ref{figure:ContourG}. 
}
\label{figure:ContourZjj}
\end{figure}

\subsubsection{Results} 
 The signal significances calculated using  Eq.~\eqref{eq:significance}  are tabulated in Table~\ref{table:SigZjj}.  We see similar behavior for the different BPs as in the previously discussed channels, i.e. enhanced signal significance with decreasing mass of the DM. The selection cuts are most efficient for SP- and TAT-type operators. For BP-$1$ we remove more than $98\%$ of the background events while keeping at least $21\%$ of the signal events for the two operator types, yielding a large signal significance especially for the TAT-type operator with polarized beams. \par
Varying the DM mass, we display the $3\sigma$ sensitivity contours for all three operators in  Fig.~\ref{figure:ContourZjj}. It is clear that the TAT-type operator has the best sensitivity, which reaches up to $4.2$~TeV with unpolarized beams and $5.2$~TeV with optimally polarized beams. The SP-type operator also has a sensitivity comparable to the mono-photon channel, and can reach up to 3.4 (4.2) TeV with unpolarized (polarized) beams. The VA-type operator has a modest sensitivity in this channel, only up to 1.7 (2.7) TeV with unpolarized (polarized) beams. 
\section{Conclusion} \label{sec:5}
We have explored the physics potential of the future $e^+e^-$ colliders in probing such {\it leptophilic} DM in a model-independent way. As a case study, we have taken the $\sqrt s=1$ TeV ILC with an integrated luminosity of 1000 fb$^{-1}$ and have analyzed the  
pair-production of fermionic DM using leptophilic dimension-6 operators of all possible bilinear structures, namely, scalar-pseudoscalar, vector-axialvector and tensor-axialtensor. We have performed a detailed cut-based analysis for each of these operators in three different channels based on the tagged particle, namely, mono-photon,  mono-$Z$ leptonic and hadronic. 

We have taken into account one of the most important and powerful features of lepton colliders, i.e., the possibility of beam-polarization with different degrees of polarization and helicity orientations. We find that the ${\rm sign}(P(e^-),P(e^+))=(+,+)$ beam configuration is optimal for the SP and TAT-type operators, while the $(+,-)$ configuration is better for probing the VA-type operators. The maximum value of the cut-off scale $\Lambda$ that can be probed in each channel at $3\sigma$ is given in Table~\ref{tab:summary}. We find that without any systematics, the mono-photon channel provides the best sensitivity across all operator types, while in presence of background systematic effects, the mono-$Z$ hadronic channel provides better sensitivity for the SP and TAT-type operators. 

We also demonstrate the complementarity of our lepton collider study with other existing direct and indirect detection searches for LDM (cf.~Figures~\ref{figure:ContourG}, \ref{figure:ContourZ} and \ref{figure:ContourZjj}). In particular, we show that lepton colliders will be able to provide the best-ever sensitivity in the still unexplored light DM regime.  
\begin{table}[t!]
    \centering
    \tbl{Summary of our results for the $3\sigma$ sensitivity reach of the cut-off scale $\Lambda$ in the three different channels discussed in the text. Here we have fixed the DM mass at 1 GeV. The numbers in parentheses are with 1\% background systematics. The numbers in bold show the highest $\Lambda$ value that can be probed for a given operator.}
 {   \begin{tabular}{|c|c|p{6em} p{6em} p{6em}|}
    \hline\hline
    {\bf Process} & {\bf Beam} & \multicolumn{3}{c|}{\bf $3\sigma$ sensitivity reach of $\Lambda$ (TeV)}\\
    \cline{3-5}
    {\bf type} & {\bf configuration} & SP & VA & TAT \\
         \hline\hline
    \multirow{2}{5em}{\centering Mono-$\gamma$} & Unpolarized & 3.91 (1.99) & 4.19 (2.14) & 4.25 (2.17) \\
     & Polarized & {\bf 4.84} (2.81) & \bf{6.49 (3.94)} & {\bf 5.28} (3.08) \\
         \hline
    Mono-$Z$ & Unpolarized & 2.62 (2.42) & 1.57 (1.36) & 3.22 (3.00) \\
    leptonic & Polarized & 3.24 (3.16) & 2.47 (2.39) & 4.02 (3.93) \\
         \hline
    Mono-$Z$ & Unpolarized & 3.38 (2.79) & 1.74 (1.39) & 4.22 (3.38) \\
    hadronic & Polarized & 4.21 ({\bf 3.87}) & 2.75 (2.57) & 5.25 ({\bf 4.71}) \\
         \hline
    \end{tabular}
    }
    \label{tab:summary}
\end{table}

\section*{Acknowledgment}
The work of BD is supported in part by the US DOE Grant \#DE-SC0017987.

\bibliographystyle{JHEP}
\bibliography{ref}

\providecommand{\href}[2]{#2}\begingroup\raggedright\begin{thebibliography}{10}

\bibitem{Krauss:2002px}
L.~M. Krauss, S.~Nasri and M.~Trodden, \emph{{A Model for neutrino masses and
  dark matter}}, \href{https://doi.org/10.1103/PhysRevD.67.085002}{\emph{Phys.
  Rev. D} {\bfseries 67} (2003) 085002}
  [\href{https://arxiv.org/abs/hep-ph/0210389}{{\ttfamily hep-ph/0210389}}].

\bibitem{Baltz:2002we}
E.~A. Baltz and L.~Bergstrom, \emph{{Detection of leptonic dark matter}},
  \href{https://doi.org/10.1103/PhysRevD.67.043516}{\emph{Phys. Rev. D}
  {\bfseries 67} (2003) 043516}
  [\href{https://arxiv.org/abs/hep-ph/0211325}{{\ttfamily hep-ph/0211325}}].

\bibitem{Ma:2006km}
E.~Ma, \emph{{Verifiable radiative seesaw mechanism of neutrino mass and dark
  matter}}, \href{https://doi.org/10.1103/PhysRevD.73.077301}{\emph{Phys. Rev.
  D} {\bfseries 73} (2006) 077301}
  [\href{https://arxiv.org/abs/hep-ph/0601225}{{\ttfamily hep-ph/0601225}}].

\bibitem{Hambye:2006zn}
T.~Hambye, K.~Kannike, E.~Ma and M.~Raidal, \emph{{Emanations of Dark Matter:
  Muon Anomalous Magnetic Moment, Radiative Neutrino Mass, and Novel
  Leptogenesis at the TeV Scale}},
  \href{https://doi.org/10.1103/PhysRevD.75.095003}{\emph{Phys. Rev. D}
  {\bfseries 75} (2007) 095003}
  [\href{https://arxiv.org/abs/hep-ph/0609228}{{\ttfamily hep-ph/0609228}}].

\bibitem{Bernabei:2007gr}
R.~Bernabei et~al., \emph{{Investigating electron interacting dark matter}},
  \href{https://doi.org/10.1103/PhysRevD.77.023506}{\emph{Phys. Rev. D}
  {\bfseries 77} (2008) 023506}
  [\href{https://arxiv.org/abs/0712.0562}{{\ttfamily 0712.0562}}].

\bibitem{Cirelli:2008pk}
M.~Cirelli, M.~Kadastik, M.~Raidal and A.~Strumia, \emph{{Model-independent
  implications of the e+-, anti-proton cosmic ray spectra on properties of Dark
  Matter}}, \href{https://doi.org/10.1016/j.nuclphysb.2008.11.031}{\emph{Nucl.
  Phys. B} {\bfseries 813} (2009) 1}
  [\href{https://arxiv.org/abs/0809.2409}{{\ttfamily 0809.2409}}].

\bibitem{Chen:2008dh}
C.-R. Chen and F.~Takahashi, \emph{{Cosmic rays from Leptonic Dark Matter}},
  \href{https://doi.org/10.1088/1475-7516/2009/02/004}{\emph{JCAP} {\bfseries
  02} (2009) 004} [\href{https://arxiv.org/abs/0810.4110}{{\ttfamily
  0810.4110}}].

\bibitem{Bi:2009md}
X.-J. Bi, P.-H. Gu, T.~Li and X.~Zhang, \emph{{ATIC and PAMELA Results on
  Cosmic e+- Excesses and Neutrino Masses}},
  \href{https://doi.org/10.1088/1126-6708/2009/04/103}{\emph{JHEP} {\bfseries
  04} (2009) 103} [\href{https://arxiv.org/abs/0901.0176}{{\ttfamily
  0901.0176}}].

\bibitem{Ibarra:2009bm}
A.~Ibarra, A.~Ringwald, D.~Tran and C.~Weniger, \emph{{Cosmic Rays from
  Leptophilic Dark Matter Decay via Kinetic Mixing}},
  \href{https://doi.org/10.1088/1475-7516/2009/08/017}{\emph{JCAP} {\bfseries
  08} (2009) 017} [\href{https://arxiv.org/abs/0903.3625}{{\ttfamily
  0903.3625}}].

\bibitem{Dev:2013hka}
P.~S.~B. Dev, D.~K. Ghosh, N.~Okada and I.~Saha, \emph{{Neutrino Mass and Dark
  Matter in light of recent AMS-02 results}},
  \href{https://doi.org/10.1103/PhysRevD.89.095001}{\emph{Phys. Rev. D}
  {\bfseries 89} (2014) 095001}
  [\href{https://arxiv.org/abs/1307.6204}{{\ttfamily 1307.6204}}].

\bibitem{Chang:2014tea}
S.~Chang, R.~Edezhath, J.~Hutchinson and M.~Luty, \emph{{Leptophilic Effective
  WIMPs}}, \href{https://doi.org/10.1103/PhysRevD.90.015011}{\emph{Phys. Rev.
  D} {\bfseries 90} (2014) 015011}
  [\href{https://arxiv.org/abs/1402.7358}{{\ttfamily 1402.7358}}].

\bibitem{Agrawal:2014ufa}
P.~Agrawal, Z.~Chacko and C.~B. Verhaaren, \emph{{Leptophilic Dark Matter and
  the Anomalous Magnetic Moment of the Muon}},
  \href{https://doi.org/10.1007/JHEP08(2014)147}{\emph{JHEP} {\bfseries 08}
  (2014) 147} [\href{https://arxiv.org/abs/1402.7369}{{\ttfamily 1402.7369}}].

\bibitem{Bell:2014tta}
N.~F. Bell, Y.~Cai, R.~K. Leane and A.~D. Medina, \emph{{Leptophilic dark
  matter with $Z'$ interactions}},
  \href{https://doi.org/10.1103/PhysRevD.90.035027}{\emph{Phys. Rev. D}
  {\bfseries 90} (2014) 035027}
  [\href{https://arxiv.org/abs/1407.3001}{{\ttfamily 1407.3001}}].

\bibitem{Freitas:2014jla}
A.~Freitas and S.~Westhoff, \emph{{Leptophilic Dark Matter in Lepton
  Interactions at LEP and ILC}},
  \href{https://doi.org/10.1007/JHEP10(2014)116}{\emph{JHEP} {\bfseries 10}
  (2014) 116} [\href{https://arxiv.org/abs/1408.1959}{{\ttfamily 1408.1959}}].

\bibitem{Cao:2014cda}
Q.-H. Cao, C.-R. Chen and T.~Gong, \emph{{Leptophilic dark matter confronts
  AMS-02 cosmic-ray positron flux}},
  \href{https://doi.org/10.1016/j.cjph.2016.11.006}{\emph{Chin. J. Phys.}
  {\bfseries 55} (2017) 10} [\href{https://arxiv.org/abs/1409.7317}{{\ttfamily
  1409.7317}}].

\bibitem{Lu:2016ups}
B.-Q. Lu and H.-S. Zong, \emph{{Leptophilic dark matter in Galactic Center
  excess}}, \href{https://doi.org/10.1103/PhysRevD.93.083504}{\emph{Phys. Rev.
  D} {\bfseries 93} (2016) 083504}.

\bibitem{Duan:2017pkq}
G.~H. Duan, L.~Feng, F.~Wang, L.~Wu, J.~M. Yang and R.~Zheng, \emph{{Simplified
  TeV leptophilic dark matter in light of DAMPE data}},
  \href{https://doi.org/10.1007/JHEP02(2018)107}{\emph{JHEP} {\bfseries 02}
  (2018) 107} [\href{https://arxiv.org/abs/1711.11012}{{\ttfamily
  1711.11012}}].

\bibitem{Madge:2018gfl}
E.~Madge and P.~Schwaller, \emph{{Leptophilic dark matter from gauged lepton
  number: Phenomenology and gravitational wave signatures}},
  \href{https://doi.org/10.1007/JHEP02(2019)048}{\emph{JHEP} {\bfseries 02}
  (2019) 048} [\href{https://arxiv.org/abs/1809.09110}{{\ttfamily
  1809.09110}}].

\bibitem{Junius:2019dci}
S.~Junius, L.~Lopez-Honorez and A.~Mariotti, \emph{{A feeble window on
  leptophilic dark matter}},
  \href{https://doi.org/10.1007/JHEP07(2019)136}{\emph{JHEP} {\bfseries 07}
  (2019) 136} [\href{https://arxiv.org/abs/1904.07513}{{\ttfamily
  1904.07513}}].

\bibitem{Ghosh:2020fdc}
S.~Ghosh, A.~Dutta~Banik, E.~J. Chun and D.~Majumdar, \emph{{Leptophilic-portal
  Dark Matter in the Light of AMS-02 positron excess}},
  \href{https://arxiv.org/abs/2003.07675}{{\ttfamily 2003.07675}}.

\bibitem{Chakraborti:2020zxt}
S.~Chakraborti and R.~Islam, \emph{{Implications of dark sector mixing on
  leptophilic scalar dark matter}},
  \href{https://doi.org/10.1007/JHEP03(2021)032}{\emph{JHEP} {\bfseries 03}
  (2021) 032} [\href{https://arxiv.org/abs/2007.13719}{{\ttfamily
  2007.13719}}].

\bibitem{Horigome:2021qof}
S.-I. Horigome, T.~Katayose, S.~Matsumoto and I.~Saha, \emph{{Leptophilic
  fermion WIMP: Role of future lepton colliders}},
  \href{https://doi.org/10.1103/PhysRevD.104.055001}{\emph{Phys. Rev. D}
  {\bfseries 104} (2021) 055001}
  [\href{https://arxiv.org/abs/2102.08645}{{\ttfamily 2102.08645}}].

\bibitem{Abi:2021ojo}
{\scshape Muon g-2} collaboration, \emph{{Measurement of the Positive Muon
  Anomalous Magnetic Moment to 0.46 ppm}},
  \href{https://doi.org/10.1103/PhysRevLett.126.141801}{\emph{Phys. Rev. Lett.}
  {\bfseries 126} (2021) 141801}
  [\href{https://arxiv.org/abs/2104.03281}{{\ttfamily 2104.03281}}].

\bibitem{Bernabei:2020mon}
{\scshape DAMA/LIBRA} collaboration, \emph{{The DAMA project: Achievements,
  implications and perspectives}},
  \href{https://doi.org/10.1016/j.ppnp.2020.103810}{\emph{Prog. Part. Nucl.
  Phys.} {\bfseries 114} (2020) 103810}.

\bibitem{Abdollahi:2017nat}
{\scshape Fermi-LAT} collaboration, \emph{{Cosmic-ray electron-positron
  spectrum from 7 GeV to 2 TeV with the Fermi Large Area Telescope}},
  \href{https://doi.org/10.1103/PhysRevD.95.082007}{\emph{Phys. Rev. D}
  {\bfseries 95} (2017) 082007}
  [\href{https://arxiv.org/abs/1704.07195}{{\ttfamily 1704.07195}}].

\bibitem{DAMPE:2017fbg}
{\scshape DAMPE} collaboration, \emph{{Direct detection of a break in the
  teraelectronvolt cosmic-ray spectrum of electrons and positrons}},
  \href{https://doi.org/10.1038/nature24475}{\emph{Nature} {\bfseries 552}
  (2017) 63} [\href{https://arxiv.org/abs/1711.10981}{{\ttfamily 1711.10981}}].

\bibitem{Adriani:2018ktz}
{\scshape CALET} collaboration, \emph{{Extended Measurement of the Cosmic-Ray
  Electron and Positron Spectrum from 11 GeV to 4.8 TeV with the Calorimetric
  Electron Telescope on the International Space Station}},
  \href{https://doi.org/10.1103/PhysRevLett.120.261102}{\emph{Phys. Rev. Lett.}
  {\bfseries 120} (2018) 261102}
  [\href{https://arxiv.org/abs/1806.09728}{{\ttfamily 1806.09728}}].

\bibitem{AMS:2021nhj}
{\scshape AMS} collaboration, \emph{{The Alpha Magnetic Spectrometer (AMS) on
  the international space station: Part II \textemdash{} Results from the first
  seven years}},
  \href{https://doi.org/10.1016/j.physrep.2020.09.003}{\emph{Phys. Rept.}
  {\bfseries 894} (2021) 1}.

\bibitem{TheFermi-LAT:2015kwa}
{\scshape Fermi-LAT} collaboration, \emph{{Fermi-LAT Observations of
  High-Energy $\gamma$-Ray Emission Toward the Galactic Center}},
  \href{https://doi.org/10.3847/0004-637X/819/1/44}{\emph{Astrophys. J.}
  {\bfseries 819} (2016) 44}
  [\href{https://arxiv.org/abs/1511.02938}{{\ttfamily 1511.02938}}].

\bibitem{XENON:2020rca}
{\scshape XENON} collaboration, \emph{{Excess electronic recoil events in
  XENON1T}}, \href{https://doi.org/10.1103/PhysRevD.102.072004}{\emph{Phys.
  Rev. D} {\bfseries 102} (2020) 072004}
  [\href{https://arxiv.org/abs/2006.09721}{{\ttfamily 2006.09721}}].

\bibitem{XENON100:2015tol}
{\scshape XENON100} collaboration, \emph{{Exclusion of Leptophilic Dark Matter
  Models using XENON100 Electronic Recoil Data}},
  \href{https://doi.org/10.1126/science.aab2069}{\emph{Science} {\bfseries 349}
  (2015) 851} [\href{https://arxiv.org/abs/1507.07747}{{\ttfamily
  1507.07747}}].

\bibitem{XENON:2019gfn}
{\scshape XENON} collaboration, \emph{{Light Dark Matter Search with Ionization
  Signals in XENON1T}},
  \href{https://doi.org/10.1103/PhysRevLett.123.251801}{\emph{Phys. Rev. Lett.}
  {\bfseries 123} (2019) 251801}
  [\href{https://arxiv.org/abs/1907.11485}{{\ttfamily 1907.11485}}].

\bibitem{LZ:2021xov}
{\scshape LZ} collaboration, \emph{{Projected sensitivities of the LUX-ZEPLIN
  (LZ) experiment to new physics via low-energy electron recoils}},
  \href{https://arxiv.org/abs/2102.11740}{{\ttfamily 2102.11740}}.

\bibitem{Chen:2018vkr}
C.-Y. Chen, J.~Kozaczuk and Y.-M. Zhong, \emph{{Exploring leptophilic dark
  matter with NA64-$\mu$}},
  \href{https://doi.org/10.1007/JHEP10(2018)154}{\emph{JHEP} {\bfseries 10}
  (2018) 154} [\href{https://arxiv.org/abs/1807.03790}{{\ttfamily
  1807.03790}}].

\bibitem{Marsicano:2018vin}
L.~Marsicano, M.~Battaglieri, A.~Celentano, R.~De~Vita and Y.-M. Zhong,
  \emph{{Probing Leptophilic Dark Sectors at Electron Beam-Dump Facilities}},
  \href{https://doi.org/10.1103/PhysRevD.98.115022}{\emph{Phys. Rev. D}
  {\bfseries 98} (2018) 115022}
  [\href{https://arxiv.org/abs/1812.03829}{{\ttfamily 1812.03829}}].

\bibitem{Kundu:2021cmo}
S.~Kundu, A.~Guha, P.~K. Das and P.~S.~B. Dev, \emph{{A model-independent
  analysis of leptophilic dark matter at future electron-positron colliders in
  the mono-photon and mono-Z channels}},
  \href{https://arxiv.org/abs/2110.06903}{{\ttfamily 2110.06903}}.

\bibitem{Kopp:2009et}
J.~Kopp, V.~Niro, T.~Schwetz and J.~Zupan, \emph{{DAMA/LIBRA and leptonically
  interacting Dark Matter}},
  \href{https://doi.org/10.1103/PhysRevD.80.083502}{\emph{Phys. Rev. D}
  {\bfseries 80} (2009) 083502}
  [\href{https://arxiv.org/abs/0907.3159}{{\ttfamily 0907.3159}}].

\bibitem{Beltran:2010ww}
M.~Beltran, D.~Hooper, E.~W. Kolb, Z.~A.~C. Krusberg and T.~M.~P. Tait,
  \emph{{Maverick dark matter at colliders}},
  \href{https://doi.org/10.1007/JHEP09(2010)037}{\emph{JHEP} {\bfseries 09}
  (2010) 037} [\href{https://arxiv.org/abs/1002.4137}{{\ttfamily 1002.4137}}].

\bibitem{Goodman:2010yf}
J.~Goodman, M.~Ibe, A.~Rajaraman, W.~Shepherd, T.~M.~P. Tait and H.-B. Yu,
  \emph{{Constraints on Light Majorana dark Matter from Colliders}},
  \href{https://doi.org/10.1016/j.physletb.2010.11.009}{\emph{Phys. Lett. B}
  {\bfseries 695} (2011) 185}
  [\href{https://arxiv.org/abs/1005.1286}{{\ttfamily 1005.1286}}].

\bibitem{Bai:2010hh}
Y.~Bai, P.~J. Fox and R.~Harnik, \emph{{The Tevatron at the Frontier of Dark
  Matter Direct Detection}},
  \href{https://doi.org/10.1007/JHEP12(2010)048}{\emph{JHEP} {\bfseries 12}
  (2010) 048} [\href{https://arxiv.org/abs/1005.3797}{{\ttfamily 1005.3797}}].

\bibitem{Goodman:2010ku}
J.~Goodman, M.~Ibe, A.~Rajaraman, W.~Shepherd, T.~M.~P. Tait and H.-B. Yu,
  \emph{{Constraints on Dark Matter from Colliders}},
  \href{https://doi.org/10.1103/PhysRevD.82.116010}{\emph{Phys. Rev. D}
  {\bfseries 82} (2010) 116010}
  [\href{https://arxiv.org/abs/1008.1783}{{\ttfamily 1008.1783}}].

\bibitem{Fox:2011fx}
P.~J. Fox, R.~Harnik, J.~Kopp and Y.~Tsai, \emph{{LEP Shines Light on Dark
  Matter}}, \href{https://doi.org/10.1103/PhysRevD.84.014028}{\emph{Phys. Rev.
  D} {\bfseries 84} (2011) 014028}
  [\href{https://arxiv.org/abs/1103.0240}{{\ttfamily 1103.0240}}].

\bibitem{Fox:2011pm}
P.~J. Fox, R.~Harnik, J.~Kopp and Y.~Tsai, \emph{{Missing Energy Signatures of
  Dark Matter at the LHC}},
  \href{https://doi.org/10.1103/PhysRevD.85.056011}{\emph{Phys. Rev. D}
  {\bfseries 85} (2012) 056011}
  [\href{https://arxiv.org/abs/1109.4398}{{\ttfamily 1109.4398}}].

\bibitem{Rajaraman:2011wf}
A.~Rajaraman, W.~Shepherd, T.~M.~P. Tait and A.~M. Wijangco, \emph{{LHC Bounds
  on Interactions of Dark Matter}},
  \href{https://doi.org/10.1103/PhysRevD.84.095013}{\emph{Phys. Rev. D}
  {\bfseries 84} (2011) 095013}
  [\href{https://arxiv.org/abs/1108.1196}{{\ttfamily 1108.1196}}].

\bibitem{Chae:2012bq}
Y.~J. Chae and M.~Perelstein, \emph{{Dark Matter Search at a Linear Collider:
  Effective Operator Approach}},
  \href{https://doi.org/10.1007/JHEP05(2013)138}{\emph{JHEP} {\bfseries 05}
  (2013) 138} [\href{https://arxiv.org/abs/1211.4008}{{\ttfamily 1211.4008}}].

\bibitem{Kahlhoefer:2017dnp}
F.~Kahlhoefer, \emph{{Review of LHC Dark Matter Searches}},
  \href{https://doi.org/10.1142/S0217751X1730006X}{\emph{Int. J. Mod. Phys. A}
  {\bfseries 32} (2017) 1730006}
  [\href{https://arxiv.org/abs/1702.02430}{{\ttfamily 1702.02430}}].

\bibitem{Penning:2017tmb}
B.~Penning, \emph{{The pursuit of dark matter at colliders\textemdash{}an
  overview}}, \href{https://doi.org/10.1088/1361-6471/aabea7}{\emph{J. Phys. G}
  {\bfseries 45} (2018) 063001}
  [\href{https://arxiv.org/abs/1712.01391}{{\ttfamily 1712.01391}}].

\bibitem{CMS:2014jvv}
{\scshape CMS} collaboration, \emph{{Search for dark matter, extra dimensions,
  and unparticles in monojet events in proton\textendash{}proton collisions at
  $\sqrt{s} = 8$ TeV}},
  \href{https://doi.org/10.1140/epjc/s10052-015-3451-4}{\emph{Eur. Phys. J. C}
  {\bfseries 75} (2015) 235} [\href{https://arxiv.org/abs/1408.3583}{{\ttfamily
  1408.3583}}].

\bibitem{ATLAS:2015qlt}
{\scshape ATLAS} collaboration, \emph{{Search for new phenomena in final states
  with an energetic jet and large missing transverse momentum in pp collisions
  at $\sqrt{s}=$8 TeV with the ATLAS detector}},
  \href{https://doi.org/10.1140/epjc/s10052-015-3517-3}{\emph{Eur. Phys. J. C}
  {\bfseries 75} (2015) 299}
  [\href{https://arxiv.org/abs/1502.01518}{{\ttfamily 1502.01518}}].

\bibitem{CMS:2017zts}
{\scshape CMS} collaboration, \emph{{Search for new physics in final states
  with an energetic jet or a hadronically decaying $W$ or $Z$ boson and
  transverse momentum imbalance at $\sqrt{s}=13\text{ }\text{ }\mathrm{TeV}$}},
  \href{https://doi.org/10.1103/PhysRevD.97.092005}{\emph{Phys. Rev. D}
  {\bfseries 97} (2018) 092005}
  [\href{https://arxiv.org/abs/1712.02345}{{\ttfamily 1712.02345}}].

\bibitem{ATLAS:2021kxv}
{\scshape ATLAS} collaboration, \emph{{Search for new phenomena in events with
  an energetic jet and missing transverse momentum in $pp$ collisions at $\sqrt
  {s}$ =13 TeV with the ATLAS detector}},
  \href{https://doi.org/10.1103/PhysRevD.103.112006}{\emph{Phys. Rev. D}
  {\bfseries 103} (2021) 112006}
  [\href{https://arxiv.org/abs/2102.10874}{{\ttfamily 2102.10874}}].

\bibitem{Cavasonza:2016qem}
L.~A. Cavasonza, H.~Gast, M.~Kr\"amer, M.~Pellen and S.~Schael,
  \emph{{Constraints on leptophilic dark matter from the AMS-02 experiment}},
  \href{https://doi.org/10.3847/1538-4357/aa624d}{\emph{Astrophys. J.}
  {\bfseries 839} (2017) 36}
  [\href{https://arxiv.org/abs/1612.06634}{{\ttfamily 1612.06634}}].

\bibitem{John:2021ugy}
I.~John and T.~Linden, \emph{{Cosmic-Ray Positrons Strongly Constrain
  Leptophilic Dark Matter}},
  \href{https://arxiv.org/abs/2107.10261}{{\ttfamily 2107.10261}}.

\bibitem{DELPHI:2003dlq}
{\scshape DELPHI} collaboration, \emph{{Photon events with missing energy in e+
  e- collisions at s**(1/2) = 130-GeV to 209-GeV}},
  \href{https://doi.org/10.1140/epjc/s2004-02051-8}{\emph{Eur. Phys. J. C}
  {\bfseries 38} (2005) 395}
  [\href{https://arxiv.org/abs/hep-ex/0406019}{{\ttfamily hep-ex/0406019}}].

\bibitem{Birkedal:2004xn}
A.~Birkedal, K.~Matchev and M.~Perelstein, \emph{{Dark matter at colliders: A
  Model independent approach}},
  \href{https://doi.org/10.1103/PhysRevD.70.077701}{\emph{Phys. Rev. D}
  {\bfseries 70} (2004) 077701}
  [\href{https://arxiv.org/abs/hep-ph/0403004}{{\ttfamily hep-ph/0403004}}].

\bibitem{Fox:2008kb}
P.~J. Fox and E.~Poppitz, \emph{{Leptophilic Dark Matter}},
  \href{https://doi.org/10.1103/PhysRevD.79.083528}{\emph{Phys. Rev. D}
  {\bfseries 79} (2009) 083528}
  [\href{https://arxiv.org/abs/0811.0399}{{\ttfamily 0811.0399}}].

\bibitem{Konar:2009ae}
P.~Konar, K.~Kong, K.~T. Matchev and M.~Perelstein, \emph{{Shedding Light on
  the Dark Sector with Direct WIMP Production}},
  \href{https://doi.org/10.1088/1367-2630/11/10/105004}{\emph{New J. Phys.}
  {\bfseries 11} (2009) 105004}
  [\href{https://arxiv.org/abs/0902.2000}{{\ttfamily 0902.2000}}].

\bibitem{Bartels:2012ex}
C.~Bartels, M.~Berggren and J.~List, \emph{{Characterising WIMPs at a future
  $e^+e^-$ Linear Collider}},
  \href{https://doi.org/10.1140/epjc/s10052-012-2213-9}{\emph{Eur. Phys. J. C}
  {\bfseries 72} (2012) 2213}
  [\href{https://arxiv.org/abs/1206.6639}{{\ttfamily 1206.6639}}].

\bibitem{Dreiner:2012xm}
H.~Dreiner, M.~Huck, M.~Kr\"amer, D.~Schmeier and J.~Tattersall,
  \emph{{Illuminating Dark Matter at the ILC}},
  \href{https://doi.org/10.1103/PhysRevD.87.075015}{\emph{Phys. Rev. D}
  {\bfseries 87} (2013) 075015}
  [\href{https://arxiv.org/abs/1211.2254}{{\ttfamily 1211.2254}}].

\bibitem{Liu:2019ogn}
Z.~Liu, Y.-H. Xu and Y.~Zhang, \emph{{Probing dark matter particles at CEPC}},
  \href{https://doi.org/10.1007/JHEP06(2019)009}{\emph{JHEP} {\bfseries 06}
  (2019) 009} [\href{https://arxiv.org/abs/1903.12114}{{\ttfamily
  1903.12114}}].

\bibitem{Habermehl:2020njb}
M.~Habermehl, M.~Berggren and J.~List, \emph{{WIMP Dark Matter at the
  International Linear Collider}},
  \href{https://doi.org/10.1103/PhysRevD.101.075053}{\emph{Phys. Rev. D}
  {\bfseries 101} (2020) 075053}
  [\href{https://arxiv.org/abs/2001.03011}{{\ttfamily 2001.03011}}].

\bibitem{Kalinowski:2021tyr}
J.~Kalinowski, W.~Kotlarski, K.~Mekala, P.~Sopicki and A.~F. Zarnecki,
  \emph{{Sensitivity of future $e^+e^-$ colliders to processes of dark matter
  production with light mediator exchange}},
  \href{https://arxiv.org/abs/2107.11194}{{\ttfamily 2107.11194}}.

\bibitem{Barman:2021hhg}
B.~Barman, S.~Bhattacharya, S.~Girmohanta and S.~Jahedi, \emph{{Catch 'em all:
  Effective Leptophilic WIMPs at the $e^+\,e^-$ Collider}},
  \href{https://arxiv.org/abs/2109.10936}{{\ttfamily 2109.10936}}.

\bibitem{Wan:2014rhl}
N.~Wan, M.~Song, G.~Li, W.-G. Ma, R.-Y. Zhang and J.-Y. Guo, \emph{{Searching
  for dark matter via mono-$Z$ boson production at the ILC}},
  \href{https://doi.org/10.1140/epjc/s10052-014-3219-2}{\emph{Eur. Phys. J. C}
  {\bfseries 74} (2014) 3219}
  [\href{https://arxiv.org/abs/1403.7921}{{\ttfamily 1403.7921}}].

\bibitem{Yu:2014ula}
Z.-H. Yu, X.-J. Bi, Q.-S. Yan and P.-F. Yin, \emph{{Dark matter searches in the
  mono-$Z$ channel at high energy $e^+e^-$ colliders}},
  \href{https://doi.org/10.1103/PhysRevD.90.055010}{\emph{Phys. Rev. D}
  {\bfseries 90} (2014) 055010}
  [\href{https://arxiv.org/abs/1404.6990}{{\ttfamily 1404.6990}}].

\bibitem{Dutta:2017ljq}
S.~Dutta, D.~Sachdeva and B.~Rawat, \emph{{Signals of Leptophilic Dark Matter
  at the ILC}},
  \href{https://doi.org/10.1140/epjc/s10052-017-5188-8}{\emph{Eur. Phys. J. C}
  {\bfseries 77} (2017) 639}
  [\href{https://arxiv.org/abs/1704.03994}{{\ttfamily 1704.03994}}].

\bibitem{Grzadkowski:2020frj}
B.~Grzadkowski, M.~Iglicki, K.~Mekala and A.~F. Zarnecki,
  \emph{{Dark-matter-spin effects at future $e^{+} e^{-}$ colliders}},
  \href{https://doi.org/10.1007/JHEP08(2020)052}{\emph{JHEP} {\bfseries 08}
  (2020) 052} [\href{https://arxiv.org/abs/2003.06719}{{\ttfamily
  2003.06719}}].

\bibitem{Bambade:2019fyw}
P.~Bambade et~al., \emph{{The International Linear Collider: A Global
  Project}},  \href{https://arxiv.org/abs/1903.01629}{{\ttfamily 1903.01629}}.

\bibitem{CLIC:2016zwp}
{\scshape CLIC, CLICdp} collaboration, \emph{{Updated baseline for a staged
  Compact Linear Collider}},
  \href{https://arxiv.org/abs/1608.07537}{{\ttfamily 1608.07537}}.

\bibitem{CEPCStudyGroup:2018ghi}
{\scshape CEPC Study Group} collaboration, \emph{{CEPC Conceptual Design
  Report: Volume 2 - Physics \& Detector}},
  \href{https://arxiv.org/abs/1811.10545}{{\ttfamily 1811.10545}}.

\bibitem{FCC:2018evy}
{\scshape FCC} collaboration, \emph{{FCC-ee: The Lepton Collider}: {Future
  Circular Collider Conceptual Design Report Volume 2}},
  \href{https://doi.org/10.1140/epjst/e2019-900045-4}{\emph{Eur. Phys. J. ST}
  {\bfseries 228} (2019) 261}.

\bibitem{Barklow:2015tja}
T.~Barklow, J.~Brau, K.~Fujii, J.~Gao, J.~List, N.~Walker et~al., \emph{{ILC
  Operating Scenarios}},  \href{https://arxiv.org/abs/1506.07830}{{\ttfamily
  1506.07830}}.

\bibitem{Delahaye:2019omf}
J.~P. Delahaye, M.~Diemoz, K.~Long, B.~Mansouli\'e, N.~Pastrone, L.~Rivkin
  et~al., \emph{{Muon Colliders}},
  \href{https://arxiv.org/abs/1901.06150}{{\ttfamily 1901.06150}}.

\bibitem{Matsumoto:2016hbs}
S.~Matsumoto, S.~Mukhopadhyay and Y.-L.~S. Tsai, \emph{{Effective Theory of
  WIMP Dark Matter supplemented by Simplified Models: Singlet-like Majorana
  fermion case}}, \href{https://doi.org/10.1103/PhysRevD.94.065034}{\emph{Phys.
  Rev. D} {\bfseries 94} (2016) 065034}
  [\href{https://arxiv.org/abs/1604.02230}{{\ttfamily 1604.02230}}].

\bibitem{Belyaev:2012qa}
A.~Belyaev, N.~D. Christensen and A.~Pukhov, \emph{{CalcHEP 3.4 for collider
  physics within and beyond the Standard Model}},
  \href{https://doi.org/10.1016/j.cpc.2013.01.014}{\emph{Comput. Phys. Commun.}
  {\bfseries 184} (2013) 1729}
  [\href{https://arxiv.org/abs/1207.6082}{{\ttfamily 1207.6082}}].

\bibitem{Alloul:2013bka}
A.~Alloul, N.~D. Christensen, C.~Degrande, C.~Duhr and B.~Fuks,
  \emph{{FeynRules 2.0 - A complete toolbox for tree-level phenomenology}},
  \href{https://doi.org/10.1016/j.cpc.2014.04.012}{\emph{Comput. Phys. Commun.}
  {\bfseries 185} (2014) 2250}
  [\href{https://arxiv.org/abs/1310.1921}{{\ttfamily 1310.1921}}].

\bibitem{Kilian:2007gr}
W.~Kilian, T.~Ohl and J.~Reuter, \emph{{WHIZARD: Simulating Multi-Particle
  Processes at LHC and ILC}},
  \href{https://doi.org/10.1140/epjc/s10052-011-1742-y}{\emph{Eur. Phys. J. C}
  {\bfseries 71} (2011) 1742}
  [\href{https://arxiv.org/abs/0708.4233}{{\ttfamily 0708.4233}}].

\bibitem{Behnke:2013lya}
H.~Abramowicz et~al., \emph{{The International Linear Collider Technical Design
  Report - Volume 4: Detectors}},
  \href{https://arxiv.org/abs/1306.6329}{{\ttfamily 1306.6329}}.

\bibitem{deFavereau:2013fsa}
{\scshape DELPHES 3} collaboration, \emph{{DELPHES 3, A modular framework for
  fast simulation of a generic collider experiment}},
  \href{https://doi.org/10.1007/JHEP02(2014)057}{\emph{JHEP} {\bfseries 02}
  (2014) 057} [\href{https://arxiv.org/abs/1307.6346}{{\ttfamily 1307.6346}}].

\bibitem{Potter:2016pgp}
C.~T. Potter, \emph{{DSiD: a Delphes Detector for ILC Physics Studies}},
  \href{https://arxiv.org/abs/1602.07748}{{\ttfamily 1602.07748}}.

\bibitem{Abramowicz:2010bg}
H.~Abramowicz et~al., \emph{{Forward Instrumentation for ILC Detectors}},
  \href{https://doi.org/10.1088/1748-0221/5/12/P12002}{\emph{JINST} {\bfseries
  5} (2010) P12002} [\href{https://arxiv.org/abs/1009.2433}{{\ttfamily
  1009.2433}}].

\bibitem{Guha:2018mli}
A.~Guha, P.~S.~B. Dev and P.~K. Das, \emph{{Model-independent Astrophysical
  Constraints on Leptophilic Dark Matter in the Framework of Tsallis
  Statistics}},
  \href{https://doi.org/10.1088/1475-7516/2019/02/032}{\emph{JCAP} {\bfseries
  02} (2019) 032} [\href{https://arxiv.org/abs/1810.00399}{{\ttfamily
  1810.00399}}].

\bibitem{Hamaide:2021hlp}
L.~Hamaide and C.~McCabe, \emph{{Fuelling the search for light dark
  matter-electron scattering}},
  \href{https://arxiv.org/abs/2110.02985}{{\ttfamily 2110.02985}}.

\bibitem{XENON:2018voc}
{\scshape XENON} collaboration, \emph{{Dark Matter Search Results from a One
  Ton-Year Exposure of XENON1T}},
  \href{https://doi.org/10.1103/PhysRevLett.121.111302}{\emph{Phys. Rev. Lett.}
  {\bfseries 121} (2018) 111302}
  [\href{https://arxiv.org/abs/1805.12562}{{\ttfamily 1805.12562}}].

\bibitem{PandaX-4T:2021bab}
{\scshape PandaX-4T} collaboration, \emph{{Dark Matter Search Results from the
  PandaX-4T Commissioning Run}},
  \href{https://arxiv.org/abs/2107.13438}{{\ttfamily 2107.13438}}.

\bibitem{Leane:2018kjk}
R.~K. Leane, T.~R. Slatyer, J.~F. Beacom and K.~C.~Y. Ng, \emph{{GeV-scale
  thermal WIMPs: Not even slightly ruled out}},
  \href{https://doi.org/10.1103/PhysRevD.98.023016}{\emph{Phys. Rev. D}
  {\bfseries 98} (2018) 023016}
  [\href{https://arxiv.org/abs/1805.10305}{{\ttfamily 1805.10305}}].

\bibitem{Steigman:2012nb}
G.~Steigman, B.~Dasgupta and J.~F. Beacom, \emph{{Precise Relic WIMP Abundance
  and its Impact on Searches for Dark Matter Annihilation}},
  \href{https://doi.org/10.1103/PhysRevD.86.023506}{\emph{Phys. Rev. D}
  {\bfseries 86} (2012) 023506}
  [\href{https://arxiv.org/abs/1204.3622}{{\ttfamily 1204.3622}}].

\bibitem{Hall:2009bx}
L.~J. Hall, K.~Jedamzik, J.~March-Russell and S.~M. West, \emph{{Freeze-In
  Production of FIMP Dark Matter}},
  \href{https://doi.org/10.1007/JHEP03(2010)080}{\emph{JHEP} {\bfseries 03}
  (2010) 080} [\href{https://arxiv.org/abs/0911.1120}{{\ttfamily 0911.1120}}].

\bibitem{Alwall:2014hca}
J.~Alwall, R.~Frederix, S.~Frixione, V.~Hirschi, F.~Maltoni, O.~Mattelaer
  et~al., \emph{{The automated computation of tree-level and next-to-leading
  order differential cross sections, and their matching to parton shower
  simulations}}, \href{https://doi.org/10.1007/JHEP07(2014)079}{\emph{JHEP}
  {\bfseries 07} (2014) 079} [\href{https://arxiv.org/abs/1405.0301}{{\ttfamily
  1405.0301}}].

\bibitem{Frixione:2007zp}
S.~Frixione, E.~Laenen, P.~Motylinski and B.~R. Webber, \emph{{Angular
  correlations of lepton pairs from vector boson and top quark decays in Monte
  Carlo simulations}},
  \href{https://doi.org/10.1088/1126-6708/2007/04/081}{\emph{JHEP} {\bfseries
  04} (2007) 081} [\href{https://arxiv.org/abs/hep-ph/0702198}{{\ttfamily
  hep-ph/0702198}}].

\bibitem{Artoisenet:2012st}
P.~Artoisenet, R.~Frederix, O.~Mattelaer and R.~Rietkerk, \emph{{Automatic
  spin-entangled decays of heavy resonances in Monte Carlo simulations}},
  \href{https://doi.org/10.1007/JHEP03(2013)015}{\emph{JHEP} {\bfseries 03}
  (2013) 015} [\href{https://arxiv.org/abs/1212.3460}{{\ttfamily 1212.3460}}].

\bibitem{Sjostrand:2014zea}
T.~Sj\"ostrand, S.~Ask, J.~R. Christiansen, R.~Corke, N.~Desai, P.~Ilten
  et~al., \emph{{An introduction to PYTHIA 8.2}},
  \href{https://doi.org/10.1016/j.cpc.2015.01.024}{\emph{Comput. Phys. Commun.}
  {\bfseries 191} (2015) 159}
  [\href{https://arxiv.org/abs/1410.3012}{{\ttfamily 1410.3012}}].

\bibitem{Cacciari:2008gp}
M.~Cacciari, G.~P. Salam and G.~Soyez, \emph{{The anti-$k_t$ jet clustering
  algorithm}}, \href{https://doi.org/10.1088/1126-6708/2008/04/063}{\emph{JHEP}
  {\bfseries 04} (2008) 063} [\href{https://arxiv.org/abs/0802.1189}{{\ttfamily
  0802.1189}}].

\bibitem{Cacciari:2011ma}
M.~Cacciari, G.~P. Salam and G.~Soyez, \emph{{FastJet User Manual}},
  \href{https://doi.org/10.1140/epjc/s10052-012-1896-2}{\emph{Eur. Phys. J. C}
  {\bfseries 72} (2012) 1896}
  [\href{https://arxiv.org/abs/1111.6097}{{\ttfamily 1111.6097}}].

\end{thebibliography}\endgroup

\end{document}